\documentclass[%
 reprint,
superscriptaddress,
 amsmath,amssymb,
 aps,
]{revtex4-2}

\usepackage{bigints}
\usepackage{comment} 
\usepackage{ulem}

\usepackage{graphicx}
\usepackage{dcolumn}
\usepackage{bm}
\usepackage{hyperref}

\usepackage{color}
\usepackage[dvipsnames]{xcolor}
\usepackage[capitalise]{cleveref}

\begin{document}


\title{Interface collisions with diffusive mass transport}

\author{Bastien Marguet}
\email{bastien.marguet@univ-lyon1.fr}
\affiliation{Institut Lumière Matière, UMR5306 Université Lyon 1-CNRS, Université de Lyon 69622 Villeurbanne, France}
\author{F. D. A. Aarão Reis}
\email{fabioaaraoreis@gmail.com}
\affiliation{Instituto de Física, Universidade Federal Fluminense, Avenida Litorânea s/n, 24210-340 Niterói RJ, Brazil.}
\author{Olivier Pierre-Louis}%
\email{olivier.pierre-louis@univ-lyon1.fr}
\affiliation{Institut Lumière Matière, UMR5306 Université Lyon 1-CNRS, Université de Lyon 69622 Villeurbanne, France}%

\date{\today}

\begin{abstract}
We report on a linear Langevin model that describes the evolution
of the roughness of two interfaces that move towards each other and are coupled by a diffusion field.
This model aims at describing the closing of the gap
between two two-dimensional material domains during growth,
and the subsequent formation of a rough grain boundary.
We assume that deposition occurs in the gap between the two domains
 and that
the growth units  diffuse and may attach to the edges of the domains.
 These units can also detach from edges, diffuse, and re-attach elsewhere.
For slow growth, the edge roughness increases monotonously
and then saturates at some equilibrium value.
For fast growth, the roughness exhibits a maximum just before 
the collision between the two interfaces,  which is followed by a minimum.
The peak of the roughness can be dominated by statistical fluctuations or 
by edge instabilities.
A phase diagram with three regimes is obtained: 
slow growth without peak, peak dominated by statistical fluctuations,
and peak dominated by instabilities. These results
reproduce the main features observed in Kinetic Monte Carlo simulations.
\end{abstract}

\maketitle



\section{Introduction}
\label{sec:introduction}

The scenario of nucleation, growth and merging of domains
is a central paradigm of non-equilibrium physics~\cite{Krapivsky2010,Livi2017}. However,
while a very large body of work has been devoted to 
nucleation and to growth, little is known about merging.
The collision of two 
interfaces that move towards each other is the elementary process that governs
merging of domains. 
Depending on the symmetries of the order parameter that describes the domains, such a collision process
might lead to the disappearance of the interfaces, as in magnetic
systems similar to the Ising model, 
or lead to the formation of a new interface, such as a grain boundary
formed by collision of two growing graphene flakes.

The existing literature on interface collisions mainly focused on domains
growing side by side~\cite{Saito1995,Derrida1991}. 
Recently, the collision of two parallel 
interfaces moving towards each other
and interacting only via
short-range interactions has been investigated~\cite{Reis2018}.
Two quantities were studied: the distribution of the collision 
times along the interface, and  of the roughness of the newly formed interface.
The asymptotic statistical properties of these two quantities were then determined exactly.

However, many interfaces exhibit long-range interactions.
One major source of long-range interactions is diffusion.
In this paper, we study interface collisions 
in diffusion-limited growth of 2D domains, with a focus on 
the formation of grain boundaries during the growth of 2D materials. 
However, beyond the study of 2D materials, we aim at developing methods
for a broad class of interface collision processes that could
also pertain to diffusion-limited growth,  such as
solidification limited by the diffusion of temperature or by the diffusion
of impurities~\cite{Saito1996,Langer1980}, and the growth of bacterial colonies
limited by the diffusion of nutrients~\cite{Beer2009}.

Our focus on the formation of grain boundaries
in 2D materials is motivated by the relevance 
of the control of grain boundary roughness for 
applications. Indeed, several material
properties such as electronic conductivity~\cite{Yazyev2010}, 
thermal conductivity~\cite{Evans2010,Merabia2014} and mechanical strength~\cite{Grantab2010}
crucially depend on the physical properties of grain boundaries.
In addition, we are also motivated by the perspective
of direct comparison of our results with experiments
during~\cite{Ogawa2012} or after~\cite{Yu2011} the collision.

Diffusion is crucial in the growth of two dimensional 
materials, where growth units are usually deposited
between 2D domains or flakes, and then have to diffuse to the edges
of the flakes where their attachment leads to growth.
Since the governing laws of diffusion have no intrinsic 
scale, the relevant scales are those imposed by the geometry of the 
diffusion region, leading to long-range interactions 
at the scale of the diffusion region.  These diffusion-limited
interactions lead to  several specific features.

A first well-known effect emerges from the observation that
the quantity of mass deposited per unit time on the substrate between the two interfaces 
is proportional to the distance between them. Hence, the growth speed,
which is proportional to the deposited mass, is proportional to the distance
between the two interfaces. As a consequence, the interfaces slow down as
they get closer to each other; this is the so-called Zeno effect~\cite{Elkinani1994}.
As opposed to the case of non-interacting interfaces, 
we therefore have a time-dependent average velocity of the interfaces.
Furthermore, the dependence of the growth velocity
on the distance between the two edges
leads to a diffusion-limited repulsion that suppresses out-of-phase edge
fluctuations and leaves only in-phase fluctuations~\cite{Misbah2010}.

A second consequence of diffusion-limited dynamics is the possibility
of deterministic morphological instabilities of the growing fronts.
These instabilities have been studied in many systems
and are referred to as the Mullins and Sekerka~\cite{Mullins1963} or Bales and Zangwill~\cite{Bales1990,Misbah2010} instabilities.
The roughness of the interfaces during the growth process
results from a combination of these deterministic morphological instabilities 
and statistical fluctuations.
The statistical fluctuations are known to exhibit 
many different regimes characterized by exponents
which account for the competition between kinetic
processes such as diffusion and attachment-detachment of growth units
at the edge~\cite{Misbah2010}.

As a summary,  collision of interfaces with 
diffusion-limited dynamics appears as a challenging problem involving time-dependent
average motion of the interfaces,  morphological instabilities,
and kinetics-dependent statistical fluctuations. 
In the following, we
model this process using linear Langevin
equations that are derived from a Burton-Carbrera-Frank-like model~\cite{Misbah2010}.
Results are compared to a Kinetic Monte Carlo  (KMC) model
which has been reported in~\cite{Reis2022}.
The Langevin model leads to features that are strikingly similar to
the results of KMC simulations:
(i) First, the interface roughness can exhibit a peak before or during the collision, 
and this peak disappears for slow growth or slow attachment-detachment kinetics.
(ii) Second, the peak of roughness is followed by a sharp decrease of the roughness,
and then by a slower relaxation of the newly formed interface towards equilibrium.
(iii) Third, we obtain a diagram that exhibits three different 
regimes depending on the incoming flux and the
attachment kinetics: noise-dominated peak, instability-dominated peak, and no peak.

In the following, we start with a description of the Langevin model in \cref{sec:Langevin_model}.
Then, we describe the results of the Langevin model in \cref{sec:Results}.
Finally, in \cref{sec:KMC}, these results are compared to KMC simulations.

\section{Langevin model}
\label{sec:Langevin_model}

In this section, we introduce a Langevin model that aims at describing the fluctuations
of the domain edges during diffusion-limited interface collisions. We start with the description
of a deterministic model that accounts for deposition, diffusion, and attachment-detachment at the domain edges.
Using this model, we first describe the dynamics of two straight edges moving towards each other.
These fronts exhibit an exponential slowing-down when they approach each other, known as the Zeno effect~\cite{Elkinani1994}.
Next we derive the equations that govern the evolution of perturbations
of these straight edges. This analysis reveals that the edges can be stable or unstable
depending on the growth conditions. Finally, we add Langevin forces that
account for  equilibrium and non-equilibrium statistical fluctuations.

\subsection{Deterministic model}
\begin{figure}[h]
    \centering
    \includegraphics[width=\linewidth]{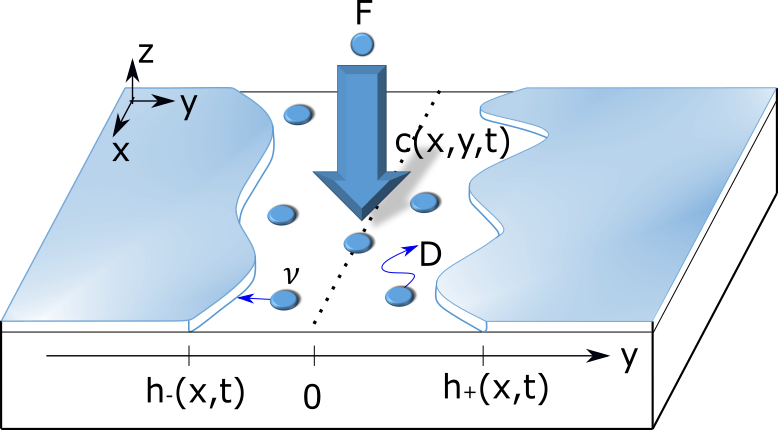}\caption{Schematic: 
    two interfaces at positions $h_+(x,t)$ and $h_-(x,t)$ grow towards each other.
    In the gap between the two interfaces,
    particles are posited with a rate $F$ per unit area and
    diffuse with a coefficient $D$. In addition, particles
    can stick to, or detach from, the edges with a kinetic coefficient $\nu$.}\label{Figure_scheme}
\end{figure}

Some key ingredients of the model are sketched in Fig.~\ref{Figure_scheme}.
Two monolayer domains grow towards each other on a substrate. 
Their edges are parallel on average.
The positions of the edges along the $y$ direction are denoted as $h_{\pm}(x,t)$. 
We assume a constant deposition flux $F$ of particles on the substrate between the two edges. 
In 2D materials such as graphene, growth units that land on top of the graphene layer
often re-evaporate quickly. As a consequence, we simply neglect them,
and attachment and detachment of growth units in the 
2D material are considered only on the substrate side. This situation
shares similarities with the Ehrlich-Schwoebel effect~\cite{Schwoebel1966,Schwoebel1969}
by which atoms attach preferentially from the lower side of atomic steps.

The areal concentration $c(x,y,t)$ of particles obeys
\begin{equation}\label{eq:non-quasistatic-diffusion}
\partial_t c(x,y,t) = D \Delta c(x,y,t) + F,
\end{equation}
where $F$ is the deposition rate and $D$ is the diffusion coefficient.
Assuming fast diffusion, we resort to the usual quasistatic approximation~\cite{Misbah2010}, 
where the concentration relaxes to a steady-state at a timescale which is much
shorter than those related to the motion of the edges. We therefore set $\partial_t c(x,y,t)=0$
in Eq.~(\ref{eq:non-quasistatic-diffusion}), leading to
\begin{equation} \label{Eqt_QS-Diffusion}
 0 = D \Delta c + F .
\end{equation}

We also assume that the properties of the edges are isotropic. The normal velocity of the edges
\begin{align}
    v_{\mathrm{n}\pm} = \mp \frac{\partial_{t} h_{\pm} (x,t)}{ [ 1+(\partial_{x}h_{\pm})^2 ]^{1/2}}
\end{align}
depends linearly on the departure from equilibrium
at the edge~\cite{Misbah2010}
\begin{equation} \label{Eqt_KBC1}
\dfrac{v_{\mathrm{n}_{\pm}}}{\Omega} = \nu (c_{\pm} - c_{\mathrm{eq},\pm}),
\end{equation}
where $\Omega$ is the specific area of a particle and $\nu$ is a kinetic coefficient that
is dimenionally homogeneous to a velocity.
Moreover, $c_{\pm}=c(x, y=\pm h(x,t),t)$ and $c_{\mathrm{eq},\pm}$ refer respectively 
to the instantaneous concentration and to the equilibrium concentration at the $\pm$ edge. 
The equilibrium concentration at the edges reads~\cite{Misbah2010}
\begin{equation}\label{Eqt_equilibrium_concentration}
c_{\mathrm{eq},\pm} = c_{\mathrm{eq}}^{0} \big( 1 +\Gamma \kappa_{\pm} \big), 
\end{equation}
where $\kappa_{\pm}$ is the curvature of the $\pm$ edge. 
The lengthscale $\Gamma=\gamma \Omega/(k_B T)$ 
is proportional to the line tension of the edge $\gamma$. 
Finally, mass conservation at the edges reads
\begin{equation}\label{Eqt_Mass_conservation1} 
\dfrac{v_{\mathrm{n}_{\pm}}}{\Omega} = \Vec{n}_{\pm} \cdot (D\Vec{\nabla}c_{\pm}),
\end{equation}
where $\Vec{n}_{\pm}$ are normal vectors that point towards the substrate side by convention. 
The system of equations (\ref{Eqt_QS-Diffusion}), (\ref{Eqt_KBC1}) and (\ref{Eqt_Mass_conservation1}) 
determines completely the dynamics of the edges.

\subsection{Straight edges}
Straight and parallel edges are a simple solution of the deterministic model.
Choosing an origin of the $y$ coordinates halfway between the two edges, we have:
\begin{equation}\label{Eqt_Def_h0+-}
h_{\pm}(x,t) = \pm {\bar{h}}^{(0)}(t) .
\end{equation}
From Eq. (\ref{Eqt_QS-Diffusion}), the associated concentration field reads
\begin{equation}\label{Eqt_Concentration_parabolic}
c^{(0)}(y,t) = {\bar{c}}^{(0)}(t) + \dfrac{F}{2D} \left( {{\bar{h}}^{(0)}}(t)^2 - y^2\right),
\end{equation} 
where the concentration at the edges is
obtained from Eqs. (\ref{Eqt_KBC1}) and (\ref{Eqt_Mass_conservation1}) as
\begin{equation}\label{Eqt_c_pm}
{\bar{c}}^{(0)}(t) = c_{\mathrm{eq}}^{0} + \dfrac{F{\bar{h}}^{(0)}(t)}{\nu}.
\end{equation}
Hence, near the edges, the concentration exceeds the equilibrium concentration 
by the ratio $F{\bar{h}}^{(0)}(t)/\nu$. This reflects the balance between 
deposition, which increases the concentration, and attachment, which decreases the concentration. 
In addition, the diffusion mass flux at the edges reads
\begin{align}\label{Eqt_massflux0}
    \mp D\partial_y c^{(0)}(y,t)=\pm F {\bar{h}}^{(0)}(t).
\end{align}
Thus, mass conservation at the edges [Eq. (\ref{Eqt_Mass_conservation1})] leads to
\begin{align}
\partial_t {\bar{h}}^{(0)} &= -\Omega F {\bar{h}}^{(0)}, \\
{\bar{h}}^{(0)}(t)&={\bar{h}}^{(0)}(0) e^{-\Omega F t}.
\label{Eqt_Interface_height}
\end{align} 
As announced above, the two edges slow down
when they approach each other. This exponential slowing-down, known as the Zeno effect\cite{Elkinani1994},
suggests that the two edges approach each other but never meet. The Zeno effect
has been invoked as the origin of the absence of merging
of mounds formed in homoepitaxial growth in the presence of a Schwoebel effect~\cite{Elkinani1994,Michely2003}.

\subsection{Deterministic dynamics of perturbations around straight edges}

The edge position is decomposed into the sum of its average $\pm {\bar{h}}^{(0)}(t)$ and of a small perturbation $h_{\pm}^{(1)}(x,t)$:
\begin{equation}\label{Eqt_PerturbationDevelopment_height_interface}
    h_{\pm}(x,t) = \pm {\bar{h}}^{(0)}(t) +  h_{\pm}^{(1)}(x,t),
\end{equation}
The same decomposition is used for the  concentration field
\begin{equation}\label{Eqt_Concentration_Development}
    c(x,y,t)=c^{(0)}(y,t)+c^{(1)}(x,y,t).
\end{equation}
In the following, we will omit the explicit dependence of $h$ and $c$ on the variables $x,y,t$ unless necessary.

From (\ref{Eqt_QS-Diffusion}), the perturbations of the concentration field obey
\begin{equation}\label{Eqt_Diffusion_1stOrder}
    \partial_{xx}c^{(1)} + \partial_{yy}c^{(1)} = 0 .
\end{equation}
The linear contribution to the boundary conditions (\ref{Eqt_KBC1},\ref{Eqt_Mass_conservation1} ) at $y=\pm {\bar{h}}^{(0)}$ 
leads to
\begin{equation}\label{Eqt_KBC_1stOrderMinusPlus}
c^{(1)}_{\pm} \pm \dfrac{D}{\nu} \partial_{y} c^{(1)}\Big|_{\pm} = \pm h_{\pm}^{(1)} \left( \dfrac{F{\bar{h}}^{(0)}}{D} + \dfrac{F}{\nu}\right) \pm c_{\mathrm{eq},\pm}^{0} \Gamma\partial_{xx} h_{\pm}^{(1)}.
\end{equation}

We define the spatial Fourier transform $f_{\mathrm{q}}$ of any function $f(x)$ as
\begin{align}
f_{\mathrm{q}} = \int^{+\infty}_{-\infty} \mathrm{d}x \; f(x) \ e^{-iqx}. 
\end{align}
Performing a Fourier transform with respect to $x$ and $t$, Eq.(\ref{Eqt_Diffusion_1stOrder}) is rewritten as
\begin{equation}\label{Eqt_Diffusion_1stOrder_Fourier}
-q^2 c_{\mathrm{q}}^{(1)}(y) + \partial_{yy}c_{\mathrm{q}}^{(1)}(y) = 0 
\end{equation}
and the concentration profile reads
\begin{equation}\label{Eqt_Diffusion_1stOrder_Fourier_Solved}
   c_{\mathrm{q}}^{(1)}(y) = a^{(1)} \operatorname{cosh} (q y) + b^{(1)} \operatorname{sinh} (q y).
\end{equation}
Using the boundary conditions (\ref{Eqt_KBC_1stOrderMinusPlus}) we obtain
\begin{align}
a^{(1)} &= \dfrac{\Delta h_{\mathrm{q}}^{(1)}}{2U_{\mathrm{q}}} \left(\dfrac{F {\bar{h}}^{(0)}}{D} + \dfrac{F}{\nu} - c_{\mathrm{eq}}^{0} \Gamma q^2 \right),  \label{Eqt_Concentration_coeffA_Fourier} \\
b^{(1)} &= \dfrac{\Sigma h_{\mathrm{q}}^{(1)}}{2V_{\mathrm{q}}} \left(\dfrac{F {\bar{h}}^{(0)}}{D} + \dfrac{F}{\nu} - c_{\mathrm{eq}}^{0} \Gamma q^2 \right), \label{Eqt_Concentration_coeffB_Fourier}
\end{align}
where we have defined the in-phase and out-of-phase modes of the edge perturbations
\begin{align}
\Sigma h_{\mathrm{q}}^{(1)} &= h_{+,q}^{(1)} + h_{-,q}^{(1)}, \label{Eqt_Def_Sigma_h}\\
\Delta h_{\mathrm{q}}^{(1)} &= h_{+,q}^{(1)} - h_{-,q}^{(1)}, \label{Eqt_Def_Delta_h}
\end{align}
and the functions of $q$
\begin{align}
U_{\mathrm{q}} = \operatorname{cosh} k + \dfrac{D}{\nu} q \operatorname{sinh} k \label{Eqt_Def_Uq}, \\
V_{\mathrm{q}} = \operatorname{sinh} k + \dfrac{D}{\nu} q \operatorname{cosh} k, \label{Eqt_Def_Vq}
\end{align}
 where $k=q {\bar{h}}^{(0)}$.
 
The deterministic dynamics of edge fluctuations are then obtained by substitution
of Eq.(\ref{Eqt_Diffusion_1stOrder_Fourier_Solved}) into  mass conservation (\ref{Eqt_Mass_conservation1}):
\begin{align}
\partial_t \Sigma h_{\mathrm{q}}^{(1)} &= \lambda_{\Sigma q} \Sigma h_{\mathrm{q}}^{(1)}, \label{Eqt_Mass_conservation_Sum_1stOrder_Fourier} \\
\partial_t \Delta h_{\mathrm{q}}^{(1)} &= \lambda_{\Delta q} \Delta h_{\mathrm{q}}^{(1)} ,\label{Eqt_Mass_conservation_Diff_1stOrder_Fourier}
\end{align}
where $\lambda_{\Sigma q}$ and $\lambda_{\Delta q}$ are the growth rates of the in-phase and out-of-phase modes
\begin{align}
\lambda_{\Sigma q} &=  \dfrac{\Omega}{V_{\mathrm{q}}} \left( F \Big( k \operatorname{cosh} k -\operatorname{sinh} k\Big) - D c_{\mathrm{eq}}^{0} \Gamma q^3 \operatorname{cosh} k \right) ,  \label{Eqt_Def_lambda_sigma_q} \\
\lambda_{\Delta q} &= \dfrac{\Omega}{U_{\mathrm{q}}} \left(  F \Big( k \operatorname{sinh} k - \operatorname{cosh} k\Big) - D c_{\mathrm{eq}}^{0} \Gamma q^3 \operatorname{sinh} k \right) . \label{Eqt_Def_lambda_delta_q}
\end{align}
A positive growth rate indicates growing perturbations, while a negative growth rate
corresponds to decaying perturbations.

These growth rates include several physical effects. 
First, the perturbations are subject to the Mullins-Sekerka instability.
Indeed, perturbations grow due to a point effect by which more atoms attach
to the protuberances of the edges. This instability is decreased
when attachment kinetics are slower (i.e. when $\nu$ is small). 
In addition, the Mullins-Sekerka instability is stronger when the incoming mass flux
in the vicinity of the edges is larger. Since these mass fluxes are proportional to ${\bar{h}}^{(0)}$ from Eq.(\ref{Eqt_massflux0}), 
the instability becomes weaker with time, as  the average distance $2{\bar{h}}^{(0)}$ between the edges decreases. 

Moreover, when the two edges are far from each other, 
perturbations from one edge are independent from those of the other edge. 
As the two edges approach each other, the perturbations of the two fronts become more and more coupled.
One important effect of this coupling
is a diffusion-limited repulsion of the two edges, leading to a strong decay of the out of phase mode.
Indeed, out-of-phase perturbations lead to a spatial variation
of the distance between the two edges. Since the edges slow down as they
approach each other, the parts where the edges are farther from each other 
grow faster, while the region where the edges are closer grow slower.
As a consequence, the perturbations of the out-of-phase mode decay.
In contrast, this mechanism does not affect in-phase perturbations of the edges
which do not lead to a spatial variation of the distance between the two edges.

Finally, line tension suppresses efficiently short wavelength perturbations of the edges.
However, long-wavelength perturbations lead to a smaller increase of the 
total length of the edges, and thus of the total energy. 
Therefore, line tension does not eliminate large wavelength fluctuations
as much as short wavelength modes.

\begin{figure}[h]
    \centering
    \includegraphics[width=0.9\columnwidth]{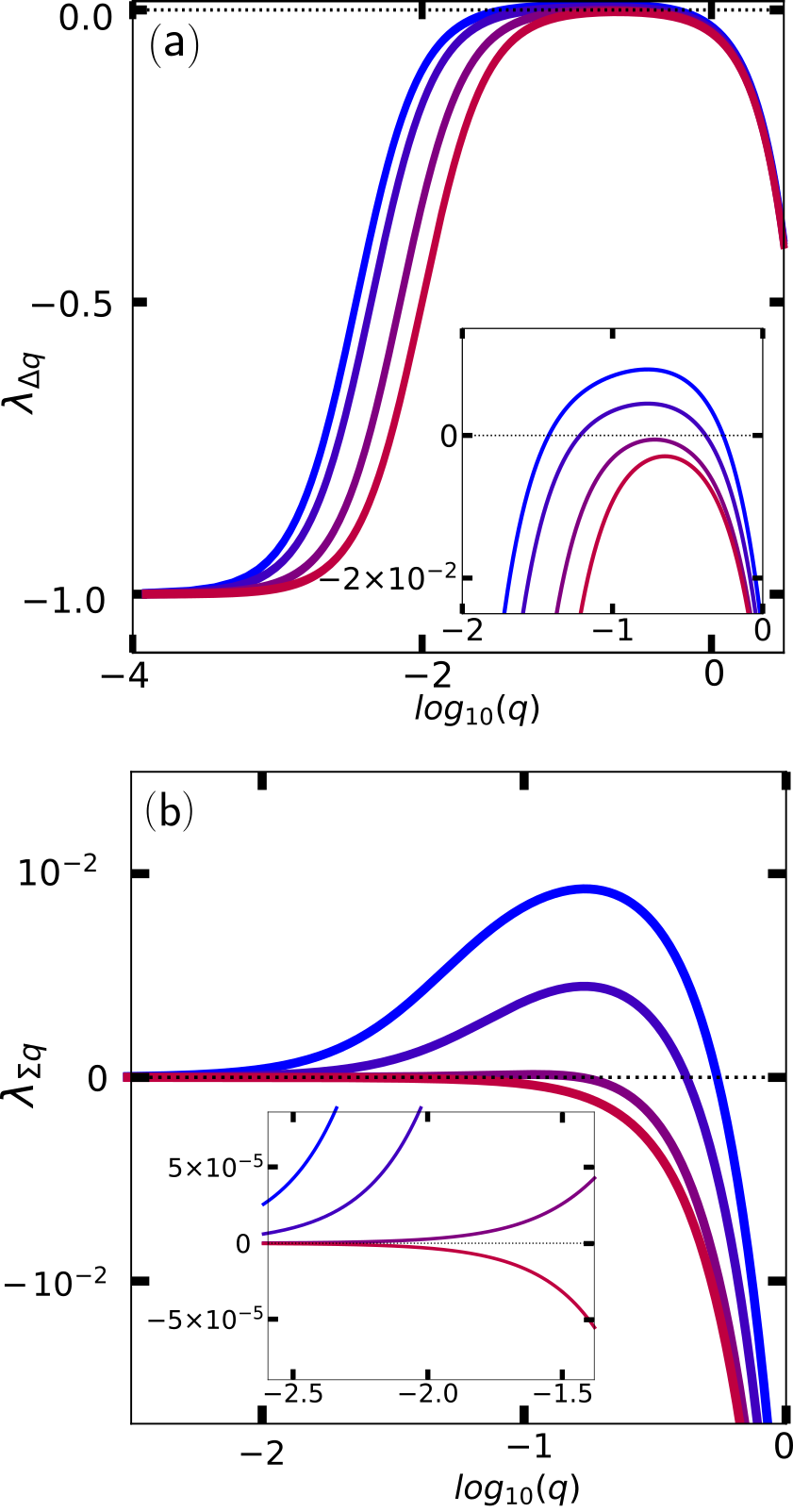}
    \caption{
    Growth rate of the perturbations. 
    (a) $\lambda_{\Sigma q}$ for the in-phase mode. 
    (b) $\lambda_{\Delta q}$ for the out-of-phase mode.
    The colors correspond to different distances between the two interfaces: 
    ${\bar{h}}^{(0)}=32$ (dark blue), ${\bar{h}}^{(0)}=20$ (blue), ${\bar{h}}^{(0)}=8$ (violet), ${\bar{h}}^{(0)}=4$ (red).
    An instability, corresponding to a positive $\lambda_{\Sigma q}$ or $\lambda_{\Delta q}$, 
    is present for large ${\bar{h}}^{(0)}$, and disappears for small ${\bar{h}}^{(0)}$.
    We have used the following model parameters:
    $\Omega=1$, $D=10^4/4$, $c_{\mathrm{eq}}^{0}=10^{-2}$, $\Gamma=4.05$, $F=1$, $\nu=1$, $L=512$.
    }
    \label{Figure_Dispersion-relation}
\end{figure}

Combining all these effects, the growth rates $\lambda_{\Sigma q}$ 
and $\lambda_{\Delta q}$ are shown on Figs.~\ref{Figure_Dispersion-relation}(a)
and \ref{Figure_Dispersion-relation}(b).
We observe that an instability, which is characterized by a 
positive value of $\lambda$, appears for both modes.
However, long-wavelength out-of-phase modes at small $q$ are always stable, 
reflecting the diffusion-limited repulsion of the two edges.
In the opposite range of the spectrum, short wavelength modes at large $q$ are always stabilized by line tension.
In addition, one observes that the instability becomes weaker with time 
and finally disappears as $\bar{h}^{(0)}$ decreases.

\subsection{Langevin description}

The deterministic growth or decay of edge perturbations
is described by \cref{Eqt_Mass_conservation_Sum_1stOrder_Fourier,Eqt_Mass_conservation_Diff_1stOrder_Fourier}. 
Nevertheless, these equations do not account for generation of the roughness due to statistical fluctuations.
We resort to Langevin forces to describe these fluctuations: 
\begin{align} 
&\partial_t \Sigma h_{\mathrm{q}}^{(1)} = \Sigma h_{\mathrm{q}}^{(1)} \  \lambda_{\Sigma q} + \eta_{\Sigma q} + \varphi_{\Sigma q}, \label{Eqt_ODE_Sum&Diff_Fourier1}\\
&\partial_t \Delta h_{\mathrm{q}}^{(1)} = \Delta h_{\mathrm{q}}^{(1)} \  \lambda_{\Delta q} + \eta_{\Delta q} + \varphi_{\Delta q}. \label{Eqt_ODE_Sum&Diff_Fourier2}
\end{align}
Langevin forces are separated into two contributions. 
The fluctuations coming from the attachment of freshly landed atoms
that have not yet been attached to an edge is accounted for by the terms $\varphi_{\Sigma q}$ and $\varphi_{\Delta q}$.
In contrast, the fluctuations that are related to the detachment-diffusion-reattachment of atoms
lead to the contributions $\eta_{\Sigma q}$ and $\eta_{\Delta q}$. 
The solution of (\ref{Eqt_ODE_Sum&Diff_Fourier1}) and (\ref{Eqt_ODE_Sum&Diff_Fourier2}) reads:
\begin{align}\label{eq:h1_Langevin}
\Sigma &h^{(1)}_{\mathrm{q}}(t) = \Sigma h^{(1)}_{\mathrm{q}}(0) \ e^{\int_0^t \mathrm{d}t' \lambda_{\Sigma q}(t') } \quad + \nonumber\\
&\int_0^t \mathrm{d}t' \left\{ e^{  \int_{t'}^{t} \mathrm{d}t'' \lambda_{\Sigma q}(t'') }\left( \eta_{\Sigma q}(t') + \varphi_{\Sigma q}(t') \right) \right\}, \\ 
\Delta &h^{(1)}_{\mathrm{q}}(t) = \Delta h^{(1)}_{\mathrm{q}}(0) \ e^{\int_0^t \mathrm{d}t' \lambda_{\Delta q}(t') } \quad + \nonumber\\
&\int_0^t \mathrm{d}t' \left\{ e^{  \int_{t'}^{t} \mathrm{d}t'' \lambda_{\Delta q}(t'') }\left( \eta_{\Delta q}(t') + \varphi_{\Delta q}(t') \right) \right\}.
\end{align}

The fluctuations of the 
each interface is characterized by its squared roughness
\begin{align}\label{Eqt_Def_roughness_Variance}
W_\pm^2(t) = \dfrac{1}{L} \int_{0}^{L} \mathrm{d}x \ h_\pm^2 - \left(\dfrac{1}{L} \int_{0}^{L} \mathrm{d}x \ h_\pm \right)^2.
\end{align}
However, it is more convenient to present our results using the squared roughness 
of the in-phase mode $\Sigma h(x,t)=h_+(x,t)+h_-(x,t)$,
which is equal to two times the average $(h_+(x,t)+h_-(x,t))/2$,
and of the out-of-phase mode $\Delta h(x,t)=h_+(x,t)-h_-(x,t)$,
which is equal to the
distance between the interfaces:
\begin{align}
 W_\Sigma^2(t)  = \dfrac{1}{L} \int_{0}^{L} \mathrm{d}x \ \Sigma h(x,t)^2 - \left(\dfrac{1}{L} \int_{0}^{L} \mathrm{d}x \ \Sigma h(x,t) \right)^2.
\label{Eqt_Def_roughness_Sigma}\\
 W_\Delta^2(t)  = \dfrac{1}{L} \int_{0}^{L} \mathrm{d}x \ \Delta h(x,t)^2 - \left(\dfrac{1}{L} \int_{0}^{L} \mathrm{d}x \ \Delta  h(x,t) \right)^2.
\label{Eqt_Def_roughness_Delta}  
\end{align}
The expected values of these squared roughnesses are
\begin{align}
\langle W_\Sigma^2(t) \rangle = \dfrac{1}{L} \int_{0}^{L} \mathrm{d}x \langle |\Sigma h^{(1)}|^2  \rangle &= \dfrac{1}{L}
\sum_{\mathrm{q}\neq 0}^{} \dfrac{\mathrm{d}q}{2\pi}  \langle |\Sigma h^{(1)}_{\mathrm{q}} |^2 \rangle, \label{Eqt_Def_roughness_WithFluctuations1}\\
\langle W_\Delta^2(t) \rangle = \dfrac{1}{L} \int_{0}^{L} \mathrm{d}x \langle |\Delta h^{(1)}|^2  \rangle &= \dfrac{1}{L} \sum_{\mathrm{q}\neq 0}^{} \dfrac{\mathrm{d}q}{2\pi}  \langle |\Delta h^{(1)}_{\mathrm{q}} |^2 \rangle,  \label{Eqt_Def_roughness_WithFluctuations2}  
\end{align}
where $\left\langle \; \right\rangle$ denotes an ensemble average 
over the fluctuations of the Langevin forces.

The correlations of the deposition noise $\varphi$ are calculated 
using a simple one-dimensional model. 
We start from the deposition
and diffusion of particles between the two edges in a discrete model,
and we take the continuum limit. The details of this procedure
are reported in Appendix~\ref{Appendix_Deposition_Noise}.
Considering a periodic system of total length $L$ along $x$, we find 
\begin{equation}\label{Eqt_Out-of-eq_Autocorrelation-function}
\langle \varphi_{i q}(t) \varphi_{j q'}(t') \rangle = 2 \Omega^2 F {\bar{h}}^{(0)} \ \delta_{i,j} \ \delta(t-t') \ \delta_{\mathrm{n+n'}}L ,
\end{equation}
where the indices $i$ and $j$ indicate either $\Sigma$ or $\Delta$.
In addition, the index $n$ accounts for Fourier modes along $x$ 
with wavenumber $q=2\pi n/L$.

As opposed to the deposition process, which is an unbalanced irreversible process
in our model, the detachment-diffusion-reattachment of atoms
can be balanced in such a way to obtain a well-defined equilibrium.
We therefore resort to a different approach based on the fluctuation-dissipation theorem to calculate the correlations
of $\eta$~\cite{PierreLouis1998}. Since this process is independent from that of 
the diffusion of freshly landed atoms, we consider an equilibrium state by simply setting
$F=0$. We then obtain from
(\ref{Eqt_ODE_Sum&Diff_Fourier1}) and (\ref{Eqt_ODE_Sum&Diff_Fourier2})
\begin{align}
 \partial_t \Sigma h_{\mathrm{q}}^{(1)} &= -\Sigma h_{\mathrm{q}}^{(1)} \ \Omega D c_{\mathrm{eq}}^{0} \Gamma q^3 \dfrac{\operatorname{cosh} k}{V_{\mathrm{q}}} + \eta_{\Sigma q}, \label{Eqt_F=0_1}\\
    \partial_t \Delta h_{\mathrm{q}}^{(1)} &= - \Delta h_{\mathrm{q}}^{(1)} \ \Omega D c_{\mathrm{eq}}^{0} \Gamma q^3 \dfrac{\operatorname{sinh} k}{U_{\mathrm{q}}}  + \eta_{\Delta q}. \label{Eqt_F=0_2}
\end{align}
Since equilibrium is a stationary process in time and the system properties are spatially homogeneous along $x$,
fluctuations are a stationary process along time $t$ and space $x$. We therefore assume that the autocorrelation of $\eta$ takes the form
\begin{align}
\langle \eta_{\Sigma q}(t) \eta_{\Sigma q'}(t') \rangle &= \delta_{\mathrm{n+n'}} \ \delta(t-t') B_{\Sigma q} L, \label{Eqt_Intensity_of_eta1}\\
\langle \eta_{\Delta q}(t) \eta_{\Delta q'}(t') \rangle &= \delta_{\mathrm{n+n'}} \ \delta(t-t') B_{\Delta q} L. \label{Eqt_Intensity_of_eta2}
\end{align}
where $B_{\Sigma q}$ and $B_{\Delta q}$ are constants. As a consequence,
the roughness of the edges  depends on $B_{\Sigma q}$ and $B_{\Delta q}$ in our model.
However, in equilibrium, the static spectrum is completely determined by the 
line stiffness $\tilde{\gamma}$ of the edge~\cite{Misbah2010}:
\begin{equation}\label{Eqt_Equilibrium_Spectrum_single_interf}
    \langle | h^{(1)}_{\mathrm{q}} |^2 \rangle_{\mathrm{eq}}=\frac{k_BT}{\tilde{\gamma} q^2}L.
\end{equation}
The roughness therefore reads:
\begin{equation}\label{Eqt_Equilibrium_Roughness}
    \langle | W^2 | \rangle_{\mathrm{eq}} 
    = \dfrac{1}{L^2} \sum_{\mathrm{n}\neq 0} \langle | h^{(1)}_{\mathrm{q}} |^2 \rangle_{\mathrm{eq}} 
    = \dfrac{k_B T L}{12 \tilde{\gamma}} = \dfrac{\Omega L}{12 \Gamma}.
\end{equation}
The consistency of this equilibrium expression with the expression of the roughness as a function of $B_{\Sigma q}$ and $B_{\Delta q}$
imposes the expression of the two amplitudes
\begin{align}
B_{\Sigma q} &= 4\Omega^2 D c_{\mathrm{eq}}^{0} \left[ \dfrac{q \operatorname{cosh} k}{V_{\mathrm{q}}}\right], \label{Eqt_Def_B_sigma_q}\\
B_{\Delta q} &= 4\Omega^2 D c_{\mathrm{eq}}^{0} \left[ \dfrac{q \operatorname{sinh} k}{U_{\mathrm{q}}}\right].\label{Eqt_Def_B_delta_q}    
\end{align}

Since the correlations of the noise are now completely determined, we obtain the expression of the 
two contributions to the time-dependent roughness from the combination of  Eqs.(\ref{eq:h1_Langevin},\ref{Eqt_Def_roughness_WithFluctuations1},\ref{Eqt_Def_roughness_WithFluctuations2})
\begin{align}
\langle W_\Sigma^2 \rangle & = \dfrac{1}{L^2} \sum_{\mathrm{n}\neq 0} 
\{ |\Sigma h^{(1)}_{\mathrm{q}}(0) |^2 \ e^{2\int_0^t \mathrm{d}t' \lambda_{\Sigma q} } \quad + \nonumber \\
&L\int_0^t \mathrm{d}t' \{ e^{ 2 \int_{t'}^{t} \mathrm{d}t'' \lambda_{\Sigma q}} ( B_{\Sigma q } + 2 \Omega^2 F {\bar{h}}^{(0)} ) \}  \}, \label{Eqt_roughness_sum_tot}
\\
\langle W_\Delta^2 \rangle & = \dfrac{1}{L^2} \sum_{\mathrm{n}\neq 0} \{ |\Delta h^{(1)}_{\mathrm{q}}(0)|^2 \ e^{2\int_0^t \mathrm{d}t' \lambda_{\Delta q} } \quad + \nonumber \\
& L\int_0^t \mathrm{d}t' \{ e^{ 2 \int_{t'}^{t} \mathrm{d}t'' \lambda_{\Delta q}} ( B_{\Delta q } + 2 \Omega^2 F {\bar{h}}^{(0)} ) \}  \}. \label{Eqt_roughness_diff_tot}
\end{align}
To evaluate these expressions numerically, we integrate their time-derivative instead of calculating directly the integrals
over time. We therefore evaluate the power spectral density $\langle |\Sigma h^{(1)}_{\mathrm{q}}(t)|^2 \rangle$ of the $\Sigma$ 
contribution to the roughness for each mode $q$ by solving
\begin{align}
\partial_t \langle |\Sigma h^{(1)}_{\mathrm{q}}(t)|^2 \rangle = 2 &\lambda_{\Sigma q}(t) \langle |\Sigma h^{(1)}_{\mathrm{q}}(t)|^2 \rangle \nonumber\\
& + B_{\Sigma q}L + 2\Omega^2 F  {\bar{h}}^{(0)}L \label{Eqt_roughness_of_modeQ_2_Sum_TimeDerived}
\end{align}
with an Euler scheme. A similar procedure is used for the $\Delta$ contribution to the roughness.

\section{Results of the Langevin Model}
\label{sec:Results}

\subsection{Temporal evolution of the roughness}

\begin{figure}[h]
    \centering
    \includegraphics[width=\linewidth]{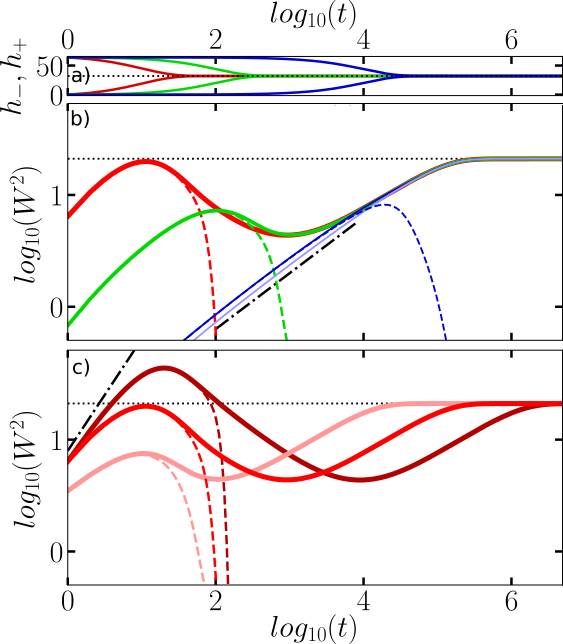}
    \caption{Average closing of the gap and evolution of the roughness. 
    a) Evolution of the interfaces average position. b) Roughness of the in-phase mode (solid curves) and out-of-phase mode (dashed curves) for $\nu=1$ and different fluxes: $F=10^{-1}$ (red), $F=10^{-2}$  (green), and $F=10^{-4}$ (blue). 
    c) Roughness of the in-phase and out-of-phase modes for $F=10^{-1}$ and different kinetic coefficients: $\nu=10^{-1}$ (dark red), $\nu=1$ (red), and $\nu=10^{1}$ (light red). Other model parameters: $\Omega=1$, $D=10^4/4$, $c_{\mathrm{eq}}^{0}=10^{-2}$, $\Gamma=4.05$, $L=512$, and $\Bar{h}^{(0)}(0)=32$. The horizontal black dotted line represents $W^2_{\mathrm{eq}}$, the squared equilibrium roughness. The dash-dotted black lines indicate the slopes associated
    to power-law behaviors: $W_{\Sigma}^2\propto t^{1/2}$ in (b) and $W_{\Sigma}^2\propto t$ in (c).
    } 
    \label{Figure_Roughness-Sigma}
\end{figure}

We have investigated the dynamics starting from straight edges at $t=0$, i.e. for all $q$
\begin{equation}\label{Eqt_Roughness_of_ModeQ_Sum&Diff_InitalTime_Discretized}
\langle |\Sigma h^{(1)}_{\mathrm{q}}(0)|^2 \rangle = \langle |\Delta h^{(1)}_{\mathrm{q}}(0)|^2 \rangle = 0 .\\
\end{equation}
Different types of evolution appear depending on the 
incoming flux $F$. They are reported in Fig.~ \ref{Figure_Roughness-Sigma}.

For small fluxes, the in-phase and out-of-phase roughnesses
are initially identical and grow due to statistical fluctuations.
Then, when the two interfaces get closer to each other,
the out-of-phase roughness decreases quickly due to
the diffusive repulsion between the two interfaces.
However, the in-phase roughness still grows
and then reaches a constant asymptotic value.

When the incoming flux is increased, the roughness 
exhibits a faster increase at short times. Once again, the 
out-of-phase roughness decreases quickly when the 
two interfaces approach each other.
However, the in-phase roughness also decreases 
when the interfaces get closer to each other.
Finally, at long times, the in-phase roughness 
starts to increase again and reaches the same asymptotic value,
which does not depend on the flux.

In the following sections, we discuss the features
of the temporal evolution of the roughnesses in more details.

\subsection{Short-time Random-Deposition roughening}
\label{s:RD_Langevin}

\begin{figure}[h]
\centering
\includegraphics[width=0.8\columnwidth]{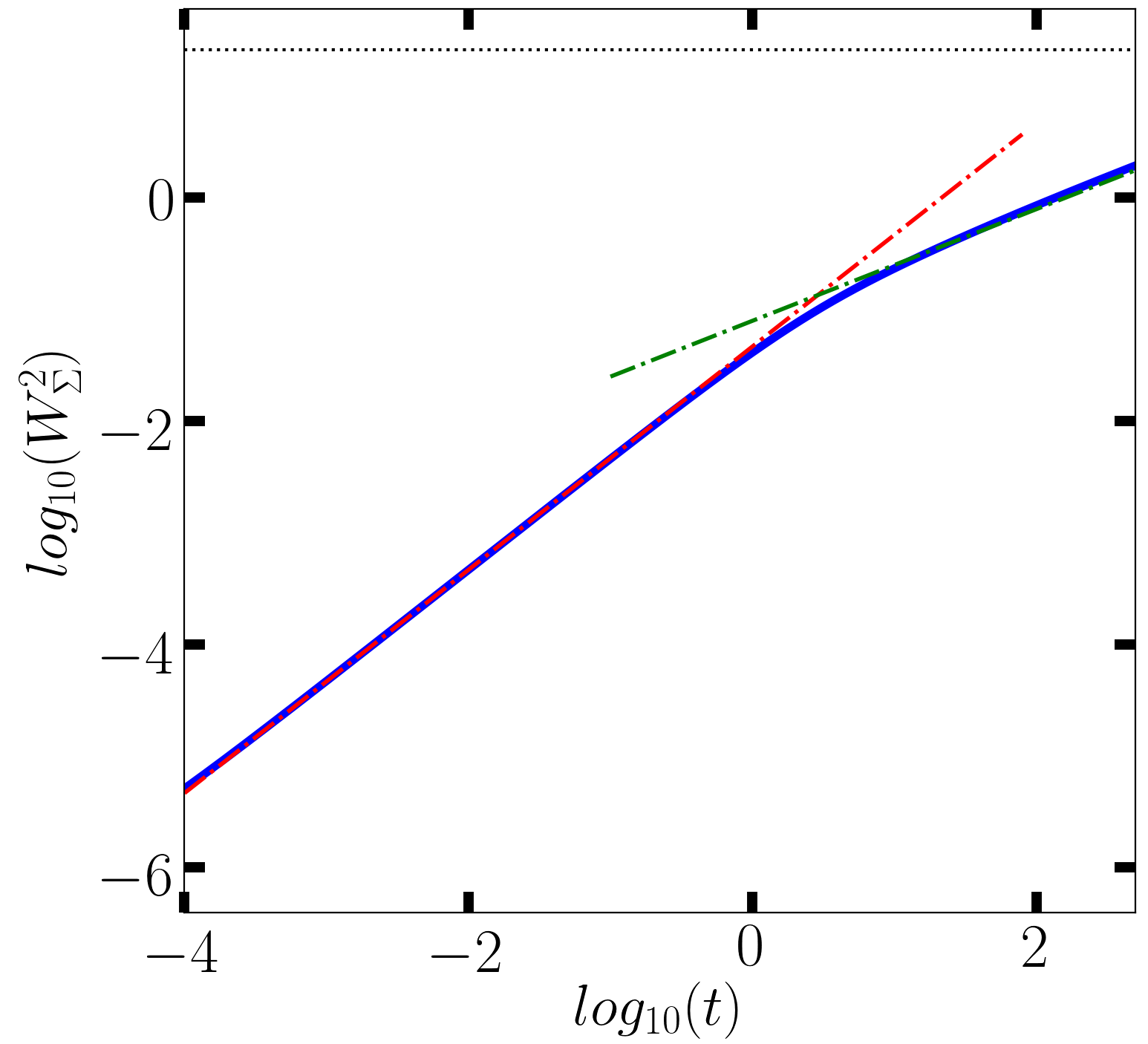}
\caption{Short-time behavior of the roughness of the in-phase mode (blue curve). 
Red dash-dotted line: asymptotic behavior $W_{\Sigma}^2\propto t$ from Eq. (\ref{eq:short_time_WSigma^2}).
Green dash-dotted line: asymptotic behavior $W_{\Sigma}^2\propto t^{1/2}$ from Eq. (\ref{eq:EW_scaling}).
The other model parameters are: $\Omega=1$, $D=10^4/4$, $c_{\mathrm{eq}}^{0}=10^{-2}$, $\Gamma=4.05$, $L=512$, ${\bar{h}}^{(0)}(0)=32$, $\nu=1$, and $F=10^{-4}$.} 
\label{Figure_Short_Time_Langevin}
\end{figure}

An expansion of Eq.(\ref{Eqt_roughness_sum_tot}) with flat initial conditions 
(\ref{Eqt_Roughness_of_ModeQ_Sum&Diff_InitalTime_Discretized}) shows that the squared roughness
is linear in time at short times:
\begin{align}
\label{eq:short_time_WSigma^2}
  \langle W_{\Sigma}^2 \rangle &= \left(
  \Omega c_{\text{eq}}^{0}\chi + \Omega F  \frac{h^{(0)}(0)}{a}
  \right) 2\Omega t,
\end{align}
where $a$ is a microscopic cutoff along $x$. The total number of modes is set to $L/2a$, which implies a cut-off for the smallest wavelength $\lambda_{\mathrm{c}}=2a$. In addition, we have defined the kinetic factor 
$\chi$, which obeys $\chi=\pi D/a^2$ for fast attachment kinetics $\nu/Dh^{(0)}(0)\gg 1$ 
and $\chi=2\nu/a$ for slow attachment kinetics $\nu/Dh^{(0)}(0)\ll 1$.
The derivation of Eq.(\ref{eq:short_time_WSigma^2}) is provided in Appendix~\ref{Appendix_Short-time_behavior}.
This linear behavior is also found in the full numerical
solution of the Langevin model, as seen in Fig.~\ref{Figure_Short_Time_Langevin}.
Other examples of this regime are reported in Fig.~\ref{Figure_Roughness_Limit_smallF_Appendix}
of Appendix~\ref{Appendix_Short-time_behavior}.

Such a linear behavior of the square roughness is associated to
the Random Deposition (RD) process, where uncorrelated attachment or detachment
events start to roughen the interface at short times~\cite{Barabasi1995,Misbah2010}.

While the first term in Eq.(\ref{eq:short_time_WSigma^2}) accounts for the 
equilibrium detachment or detachment-diffusion-reattachment events, the second term accounts for deposition-diffusion-attachment events.
Note also that $\langle W_{\Sigma}^2 \rangle$ at short times depends on
the microscopic cutoff $a$ along the $x$ axis.

\subsection{Asymptotic equilibrium roughness}

In the opposite limit of long times, 
the in-phase roughness reaches a constant value,
while the out-of-phase  modes are  efficiently eliminated
by the diffusive repulsion between the two interfaces, so that
$\langle W^2_\Delta\rangle\rightarrow 0$.
Since the amplitude of the out-of-phase modes vanishes,
the interface is composed of the two in-phase edges,
which have identical profiles at long times.
As a consequence, this effective interface with the 
two edges exhibits fluctuations that are identical to
that of an interface at equilibrium with a line stiffness $2\tilde\gamma$.
The profile of the effective interface is $(h_++h_-)/2=\Sigma h/2$,
and as a consequence its roughness is $\langle W_{\Sigma}^2 \rangle/4$.
Using the equilibrium formula [Eq.(\ref{Eqt_Equilibrium_Roughness})] for the effective interface with doubled line tension, we 
obtain $\langle W_{\Sigma}^2 \rangle/4=\Omega L/(12\times 2\Gamma)$.
This leads to the asymptotic roughnesses
\begin{align}
\label{eq:asymptotic_WSigma^2}
  \langle W_{\Sigma}^2 \rangle &= \frac{\Omega L}{6\Gamma}\,,
  \\
  \label{eq:asymptotic_WDelta^2}
  \langle W_{\Delta}^2 \rangle &\rightarrow 0\, .
\end{align}
As seen in Fig.~\ref{Figure_Roughness-Sigma}, this result is
in quantitative agreement with the asymptotic value of $\langle W_{\Sigma}^2 \rangle$ in
the full numerical solution of the Langevin model.

\subsection{Close to equilibrium Edwards-Wilkinson roughening}
\label{s:EW_Langevin}

In the limit of small fluxes $F$ [blue curve in Fig~\ref{Figure_Roughness-Sigma}(b)],
the roughness slowly builds up and increases
monotonically up to its equilibrium value (\ref{eq:asymptotic_WSigma^2}).
Since the power-spectrum at equilibrium (\ref{Eqt_Equilibrium_Spectrum_single_interf})
is dominated by long wavelengths, i.e. small $q$,
we expect that a long-wavelength expansion can catch
the main features of the convergence towards equilibrium.
In the limit $q\rightarrow 0$, a relatively simple expression of  $\langle W_\Sigma^2 \rangle$ 
can be obtained
\begin{align}
\label{eq:WSigma^2_v}
    \langle W_{\Sigma}^2 \rangle &= \frac{\Omega L}{2\pi^2 \Gamma}\sum_{\mathrm{n}\neq 0}^{} \frac{1}{n^2}(1 - \text{e}^{-v n^2}),
\end{align}
where
\begin{align}
 v= 2\Omega \Gamma \left(\frac{2\pi}{L}\right)^2\frac{\nu c_{\text{eq}}^{0}}{\Omega F}
 \ln\frac{1 + \frac{D}{\nu\Bar{h}^{(0)}(t) }}{1 + \frac{D}{\nu\Bar{h}^{(0)}(0) }} >0.
\end{align}
The derivation of this relation, reported in Appendix~\ref{a:EW},
relies on the assumption that $\lambda_{\Sigma q}$ and the noise in 
Eq.(\ref{Eqt_roughness_of_modeQ_2_Sum_TimeDerived}) can be approximated
by their values at zero flux. The effect of the flux is then only to drive
the average motion of the edges.

From Eq.(\ref{eq:WSigma^2_v}), we find that
the roughness tends to the equilibrium value of Eq.(\ref{eq:asymptotic_WSigma^2})
at long times. Indeed, when $t\rightarrow +\infty$, we have $\Bar{h}^{(0)}(t)\rightarrow 0$,
so that $v\rightarrow +\infty$ and $\text{e}^{-v n^2}\rightarrow 0$ in Eq.(\ref{eq:WSigma^2_v}),
leading to Eq.(\ref{eq:asymptotic_WSigma^2}).

In the regimes where the time $t$ is not too short, so that long wavelength 
modes have enough time to develop, {\color{red} but is also} not too large, so that the equilibrium
roughness is not reached,
we obtain the usual Edwards-Wilkinson scaling~\cite{Barabasi1995}
\begin{align}
\label{eq:EW_scaling}
  \langle W_{\Sigma}^2 \rangle &\simeq \Omega\bigg(
  \frac{\Omega c_{\text{eq}}^{(0)}}{1/\nu+\bar{h}^{(0)}(0)/D }\;
  \frac{8t}{\pi\Gamma }
  \bigg)^{1/2} ,
\end{align}
which is associated to close-to-equilibrium roughening~\cite{Barabasi1995,Saito2012}.
The derivation of this expression is reported in Appendix~\ref{a:EW}. 

\begin{figure}[h]
\centering
\includegraphics[width=0.8\linewidth]{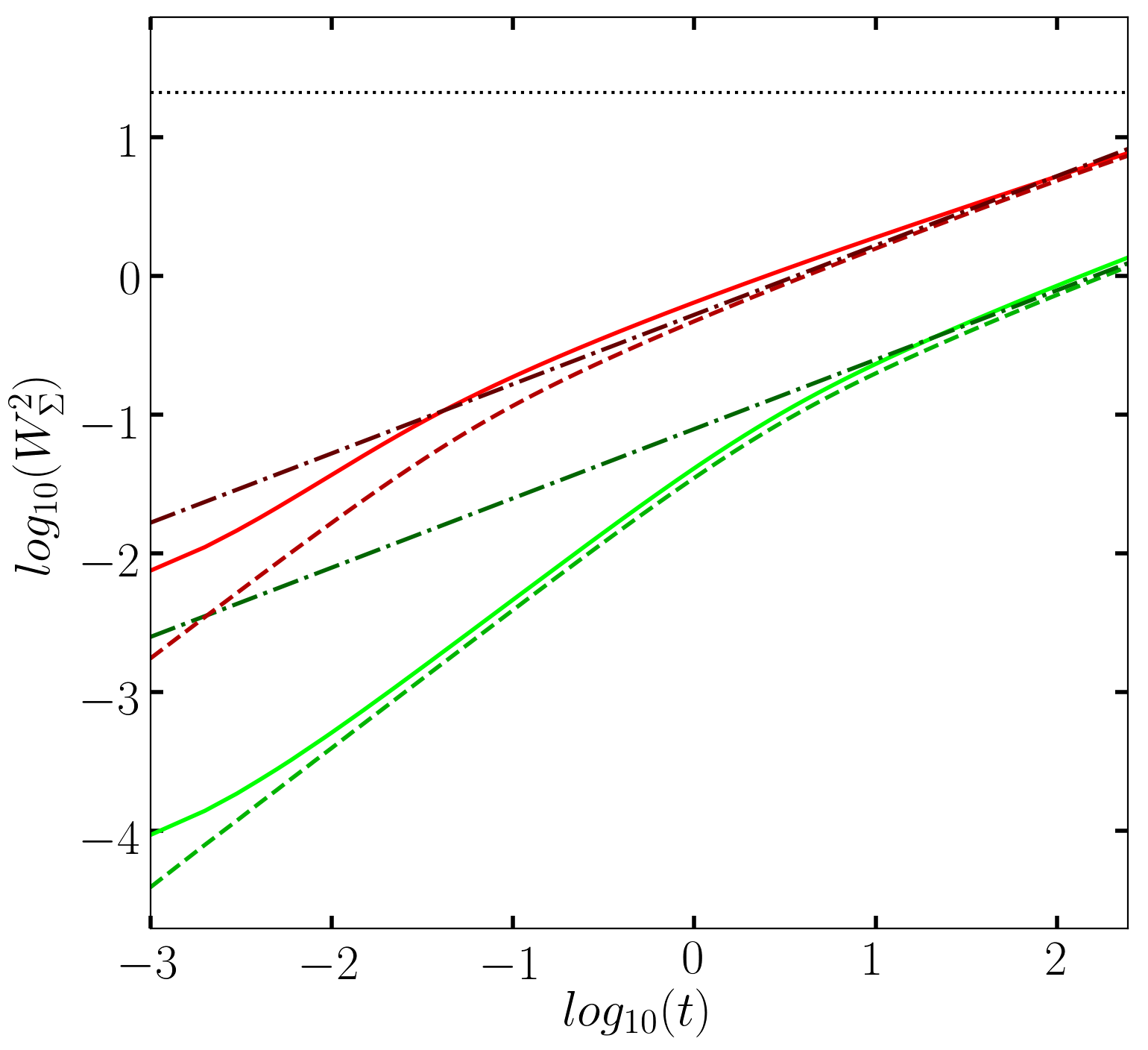}
\caption{Roughness at short and intermediate times for
slow deposition fluxes.
Solid curves: full numerical solution of the model \cref{Eqt_roughness_sum_tot}; 
Dashed curves: long-wavelength approximation \cref{eq:WSigma^2_v}; 
Dash-dotted curves: Edwards-Wilkinson scaling \cref{eq:EW_scaling}. 
Kinetic coefficients: $\nu =10^{2}$ (red) and $\nu =1$ (green). 
The other model parameters are:
$\Omega=1$, $D=10^4/4$, $c_{\mathrm{eq}}^{0}=10^{-2}$, $\Gamma=4.05$, $L=512$, ${\bar{h}}^{(0)}(0)=32$, and $F=10^{-4}$.} \label{Figure_Roughness_Limit_smallF}
\end{figure}

This expression
is seen to provide a fair approximation of the 
evolution of $\langle W_\Sigma^2 \rangle$
for small deposition fluxes $F$ and slow attachment-detachment kinetics 
as seen from Fig.~\ref{Figure_Short_Time_Langevin}.
However, note that the expressions Eq.(\ref{eq:WSigma^2_v},\ref{eq:EW_scaling})
become less accurate as attachment-detachment kinetics
becomes faster, as seen from Fig.~\ref{Figure_Roughness_Limit_smallF}.
Indeed, in the regime of fast attachment-detachment kinetics,
one cannot neglect the dependence 
of the perturbation growth rate $\lambda_{\Sigma q}$
in the deposition rate $F$.

\subsection{Peak of roughness for fast growth}

During faster growth, the initial increase of the roughness is dominated
by non-equilibrium effects, which consist of two contributions: 
(i) non-equilibrium fluctuations associated to the deposition noise $\varphi$
and (ii) the deterministic Mullins-Sekerka instability corresponding to
positive $\lambda_{\Sigma q}$ or $\lambda_{\Delta q}$.
These contributions produce roughness at short and finite wavelengths.
However, such wavelengths are strongly suppressed during the collision.
Indeed, 
growth is then slower in the late stages of the dynamics, and the line stiffness 
drives the system towards an equilibrium state
where the power-spectrum [Eq.(\ref{Eqt_Equilibrium_Spectrum_single_interf})]
is dominated by long wavelength modes.
As a consequence of the suppression of the short wavelength
modes during the collision, the roughness decreases sharply,
giving rise to a maximum of roughness for fast growth, as reported
in Fig.~\ref{Figure_Roughness-Sigma}.

Estimates of the time $t_c$ at which the peak of roughness
occur can be obtained in the limits of slow and fast attachment-detachment
kinetics. 

When the attachment-detachment kinetics is slow,
the roughening is dominated by statistical fluctuations.
Neglecting the contribution related to equilibrium
fluctuations in the right-hand-side of Eq.(\ref{Eqt_roughness_of_modeQ_2_Sum_TimeDerived})
and considering that the roughness is dominated 
by short-wavelength modes around the value of the microscopic cutoff $q_c=\pi/a$,
we find that
\begin{align}
t_{\mathrm{peak}} 
= \frac{1}{\Omega F-2 \nu \Gamma c_{\mathrm{eq}}^{0}\pi^2}
\ln \left(\frac{\Omega F }{ 2 \nu \Gamma c_{\mathrm{eq}}^{0}\pi^2}\right).
\label{eq:tc_attachment_detachment}
\end{align}
The details of this calculation are reported in Appendix~\ref{Appendix_Maximum}.

In contrast, for the regime of fast
attachment-detachment kinetics, we assume that the roughening is dominated
by the Mullins-Sekerka instability.
We therefore associate $t_c$ to the last time where the
instability is present. As discussed above, the instability 
disappears when the distance between the two interfaces decreases.
In Fig.~\ref{Figure_Dispersion-relation}, we see that
the instability disappears when the growth rate of the sum-mode $\lambda_{\Sigma q}$
changes sign at long-wavelength (i.e., small $q$).
From an expansion of $\lambda_{\Sigma q}$ at $q\rightarrow 0$,
we find that the instability disppears when ${\bar{h}}^{(0)}(t)$
becomes smaller than ${\bar{h}}^{(0)}(t_{\mathrm{c}}) = (3c_{\mathrm{eq}}^{0} \Gamma D /F)^{1/3}$.
Using Eq.(\ref{Eqt_Interface_height}), this corresponds to 
\begin{align}
    t_{\mathrm{c}}=\frac{1}{3\Omega F} \ln\Bigl(\frac{F h^{(0)\,3}_{0}}{3 c_{\mathrm{eq}}^{0}D\Gamma} \Bigr)\, .
    \label{eq:tc_instab}
\end{align}

\begin{figure}[h]
    \centering
    \includegraphics[width=\linewidth]{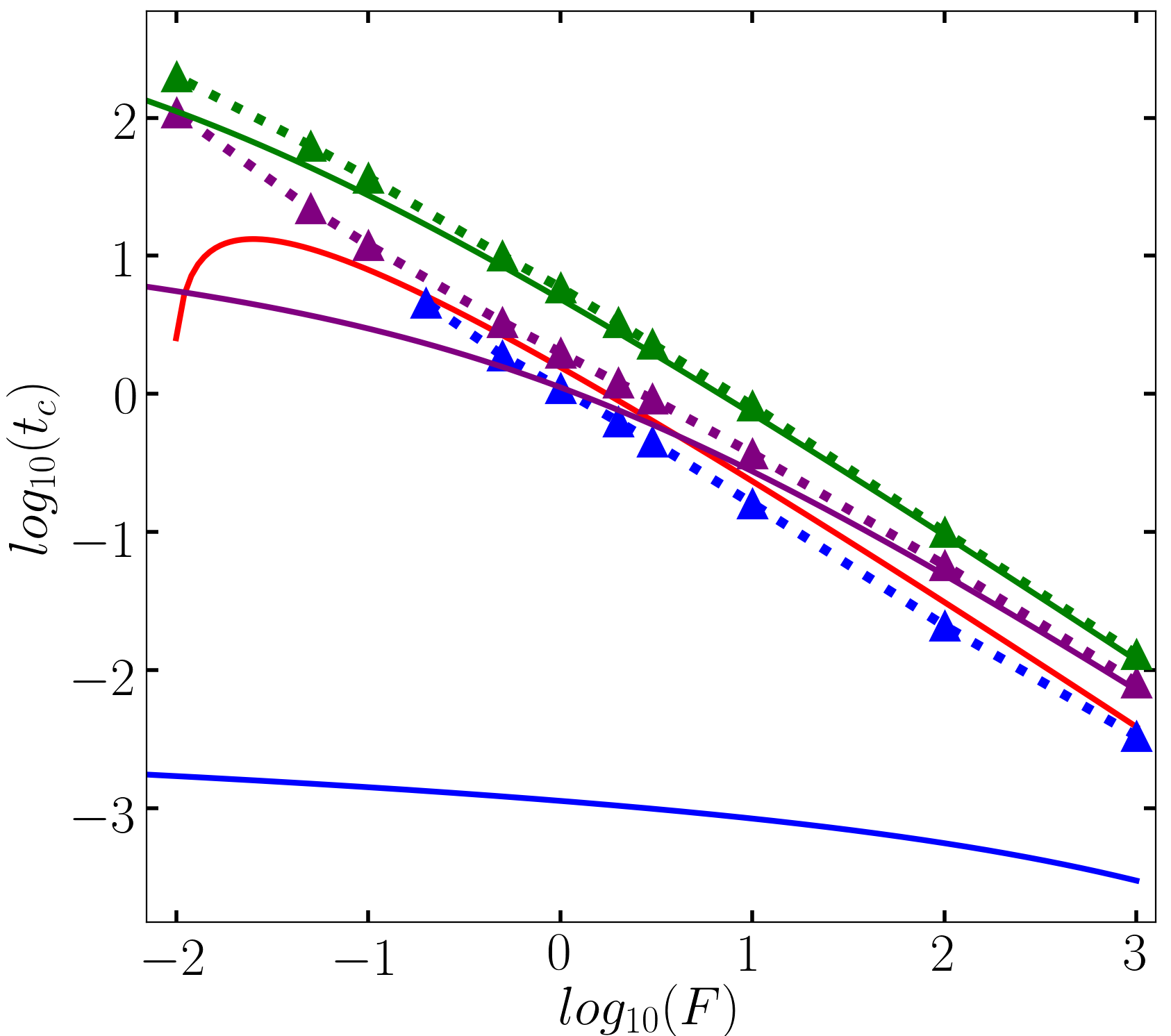}
    \caption{Time of the peak of roughness 
    as a function of the flux $F$. 
    The triangles (joined by dotted lines) are extracted from the numerical solution of the full model:
    $\nu=10^{-2}$ (green), $\nu=1$ (violet), and $\nu=10^{4}$ (blue).
    Solid curves with the same colors correspond to approximate expressions suitable for the fluctuation dominated regime
    Eq.(\ref{eq:tc_attachment_detachment}). The red solid curve reports the approximate expression Eq.(\ref{eq:tc_instab})
    for the instability-dominated regime. 
    The other model parameters are:
    $\Omega=1$, $D=10^4/4$, $c_{\mathrm{eq}}^{0}=10^{-2}$, $\Gamma=4.05$, $L=512$, and ${\bar{h}}^{(0)}(0)=32$.
    }
    \label{Figure_peak-time_flux}
\end{figure}

As reported in Fig.~\ref{Figure_peak-time_flux}, the values of the time of the peak $t_c$ obtained from the full
numerical simulations of the Langevin model at large deposition flux $F$
are in good agreement with Eq.(\ref{eq:tc_attachment_detachment})
for slow attachment-detachment kinetics,
and with Eq.(\ref{eq:tc_instab}) for fast attachment-detachment kinetics.

\subsection{Classification of the roughening regimes}
\label{s:classification_Langevin}

In this section, we wish to identify the different 
regimes for the evolution of the roughness as a function
of relevant physical parameters.
In order to identify the contributions which dominate 
the evolution of the roughness, we analyse the evolution of the ratio of the deterministic term over the stochastic one in Eq. (\ref{Eqt_roughness_of_modeQ_2_Sum_TimeDerived}): 
\begin{equation}\label{Eqt_Ratio-Sigma}
R_{\Sigma}(t) =\frac{\sum_{\mathrm{n}\neq 0}^{}2 \lambda_{\Sigma q}(t) \langle |\Sigma h^{(1)}_{\mathrm{q}}(t)|^2 \rangle}{\sum_{\mathrm{n}\neq 0}^{}\left( B_{\Sigma q}(t) + 2\Omega^2 F  {\bar{h}}^{(0)}(t)\right)L}. 
\end{equation} 
where we recall that $q=2\pi n/L$ and $L$ is the system size along $x$.
The denominator of $R_{\Sigma}$ is always positive. Thus, the sign of the numerator dictates the sign of $R_{\Sigma}$. 
The special value $R_{\Sigma}=-1$
corresponds to $\partial_t \langle |\Sigma h^{(1)}_{\mathrm{q}}(t)|^2 \rangle =0$ from Eq.~(\ref{Eqt_roughness_of_modeQ_2_Sum_TimeDerived}).
The equilibrium state which is always obtained at long times obeys this condition.
Therefore, $R_{\Sigma}$ always converge to $-1$ at long times, i.e., when $t\rightarrow\infty$.
When $R_{\Sigma}$ crosses the line $R_{\Sigma}=-1$ at finite times, the roughness reaches an extremum, 
which is either a maximum or a minimum of $W^2_{\Sigma}$.
In addition, $R_{\Sigma}$ can be positive only if $\lambda_{\Sigma q}(t)>0$
for some value of $q$, i.e., only in the presence of the Mullins-Sekerka instability.

We define three regimes corresponding
to the different types of dynamics resulting from the Langevin model.
First, the monotonic  roughening regime is observed
in the absence of deterministic Mullins-Sekerka instability. We therefore have $\lambda_{\Sigma q}(t)<0$
for all $q$ and as a consequence $R_{\Sigma}<0$. Hence, 
if $0<R_{\Sigma}<-1$ at all times, then the dynamics is considered
to be in the monotonic  roughening regime.
Second,
the instability-dominated peak regime is defined as the regime
where $R_{\Sigma}>1$ at some point during the dynamics,
i.e. when the instability contribution is larger than the noise contribution in Eq.(\ref{Eqt_roughness_of_modeQ_2_Sum_TimeDerived}). 
Third, when $R_{\Sigma}<1$ and $R_{\Sigma}$ crosses the line at $R_{\Sigma}=-1$,
we consider that the dynamics belong to the fluctuation-dominated peak regime.

\begin{figure}[h]
    \centering
    \includegraphics[width=\linewidth]{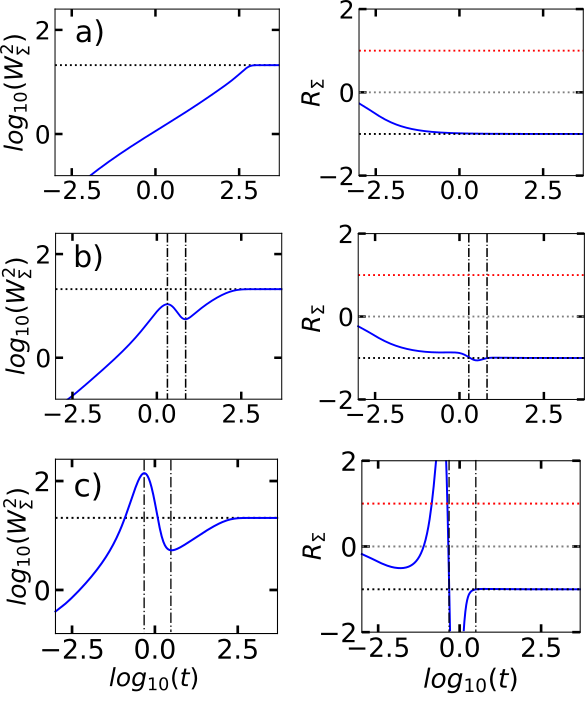}\caption{Time evolution of $W^2_{\Sigma}$ and $R_{\Sigma}$.
    $R_{\Sigma}$ is defined in (\ref{Eqt_Ratio-Sigma}) as the amplitude of the ratio of the deterministic part over the stochastic one. It gives a criterion for the domain classification in the phase diagram Fig. \ref{Figure_Stability_Diagram_with_shapes}. The vertical dash-dotted lines mark the local extrema of the roughness. They coincide with the events $R_{\Sigma}=-1$. We have used the following model parameters:
    $\nu/D=0.4$, $\Omega=1$, $D=10^4/4$, $c_{\mathrm{eq}}^{0}=10^{-2}$, $\Gamma=4.05$, $L=512$, ${\bar{h}}^{(0)}(0)=32$. 
    a) $F/D=1.2\ 10^{-3}$. b) $F/D=2.0\ 10^{-4}$. c) $F/D=2.0\ 10^{-6}$.
    }
    \label{Figure_Roughness-Sigma-criterion}
    \end{figure}

Some examples of dynamics in these three different regimes
are reported in Fig.~\ref{Figure_Roughness-Sigma-criterion}. 
The vertical dash-dotted lines mark the local extrema of the roughness. They coincide with the condition $R_{\Sigma}=-1$.

\begin{figure}[h]
    \centering
    \includegraphics[width=\linewidth]{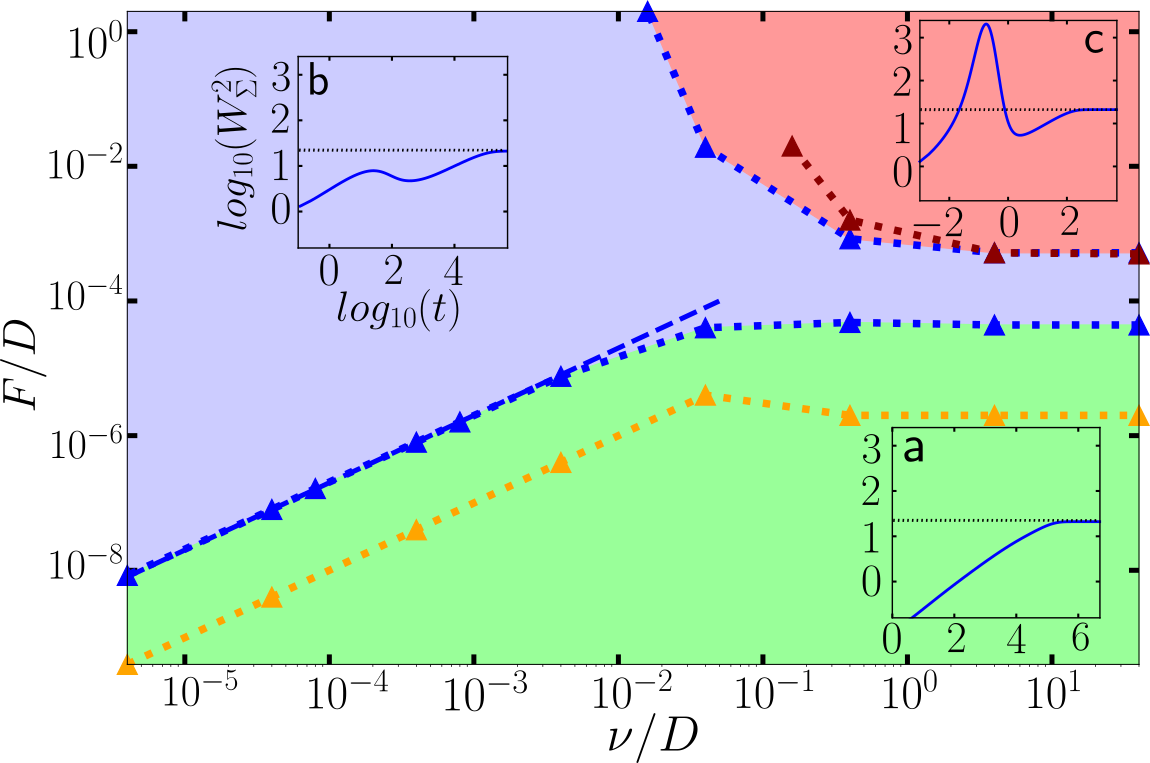}
    \caption{
    Phase diagram reporting the three main regimes in the ($F/D$, $\nu/D$) plane.
    The blue triangles define the boundaries of the three regions. 
    The green region corresponds to the monotonic roughening regime. 
    The blue region corresponds to the presence of a peak of roughness dominated by out-of-equilibrium fluctuations. 
    The red region corresponds to the presence of a peak of roughness due to a morphological instability. 
    The orange and red triangles respectively correspond to the regions
    where $W^2\propto t^{1/2}$ and where $W^2$ grows faster than $t^{2}$. 
    The parameters of the insets are: 
    a) $F/D=4.10^{-8}$, $\nu/D=4.10^{-4}$; 
    b) $F/D=4.10^{-5}$, $\nu/D=4.10^{-3}$; 
    c) $F/D=4.10^{-3}$, $\nu/D=4.10^{-1}$.
    The other model parameters are:
    $\Omega=1$, $D=10^4/4$, $c_{\mathrm{eq}}^{0}=10^{-2}$, 
    $\Gamma=4.05$, $L=512$, ${\bar{h}}^{(0)}(0)=32$.}
    \label{Figure_Stability_Diagram_with_shapes}
\end{figure}

The occurrence of these regimes as a function of $F/D$ and $\nu/D$
is summarized in Fig.~\ref{Figure_Stability_Diagram_with_shapes}.
We observe that the boundary between the monotonous roughening regime and the fluctuation-dominated peak regime 
for low attachment/detachment kinetics is linear at small $F/D$ and small $\nu/D$,
while the boundary between the monotonous roughening regime and 
the instability-dominated peak regime corresponds to a constant $F/D$.

\section{Kinetic Monte-Carlo simulations}
\label{sec:KMC}

\begin{figure}[h]
    \centering
\includegraphics[width=\linewidth]{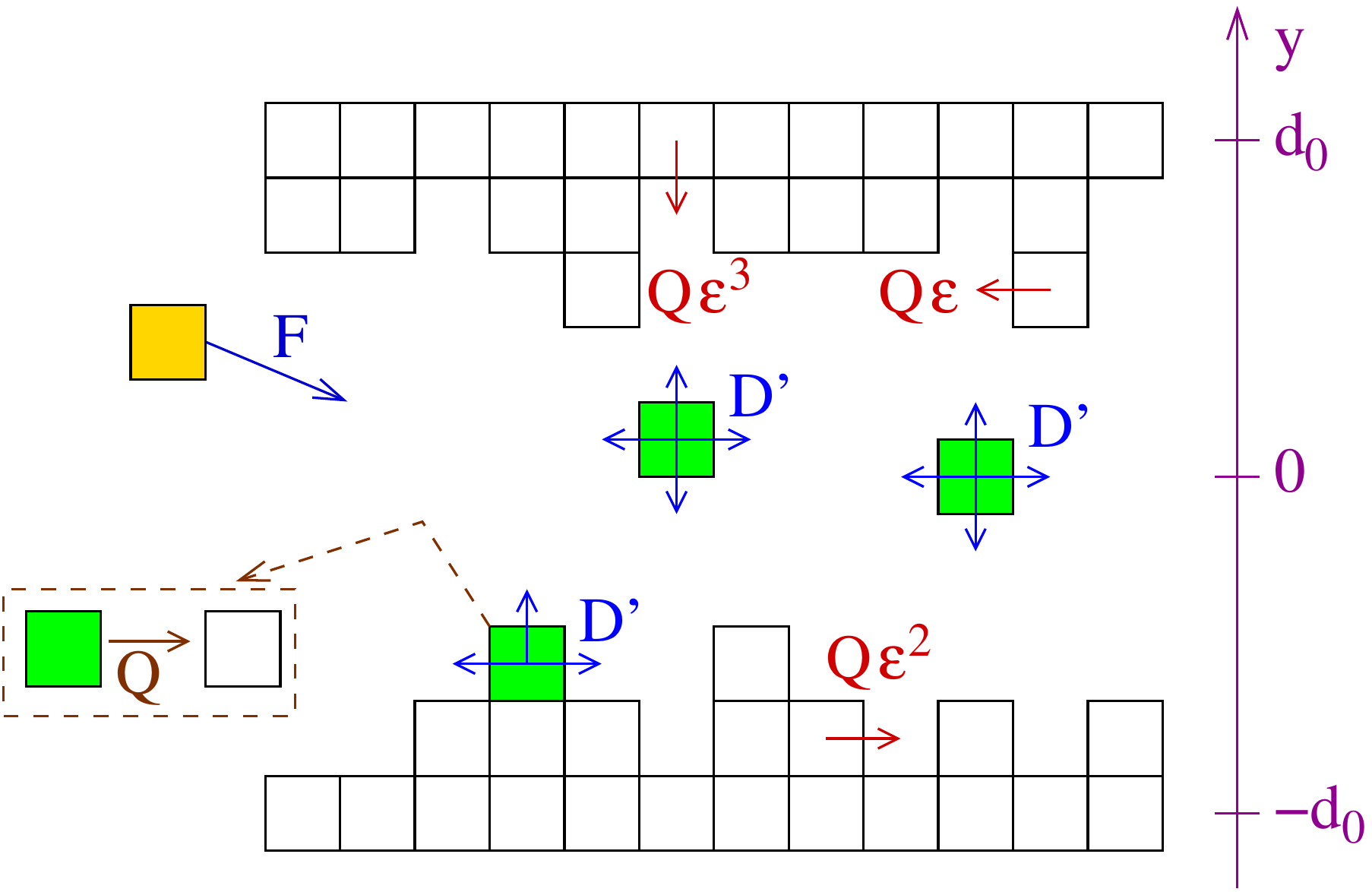}    
\caption{
    Schematic of the KMC model. Solid particles are in white, 
    mobile particles in green, and an incident particle is shown in yellow. 
    Blue arrows indicate deposition events and the possible hops of mobile particles. 
    Red arrows indicate detachment of solid particles. 
    The brown arrow shows the transformation of a mobile particle 
    into a solid particle at its current position. 
    The corresponding rates are indicated near the arrows.
    }\label{Figure_Kinetic_Monte-Carlo}
\end{figure}

\subsection{KMC model}

In this section, we present  the lattice KMC model that is discussed 
in details in Ref.~\cite{Reis2022}, and compare the simulation results
with the Langevin model discussed above.
The model is defined in a square lattice with two solids with initially flat interfaces located at $y = -d_0$ (interface $-$) and $y = +d_0$ (interface $+$). The length of each interface is L and periodic boundaries are considered in the x direction. Each atom of the solid is represented by a solid site (also called solid particle). The model is represented in Fig \ref{Figure_Kinetic_Monte-Carlo}.
We use the lattice parameter $a$ as the unit length.
As a consequence, the atomic area is $\Omega=a^2=1$.

Deposition events occur only in the region between the two solid interfaces.
The external particle flux is F, measured in number of incident particles per site per unit time. 
If an incident particle is deposited on a free site, 
it becomes a mobile particle; otherwise, the deposition attempt is rejected. 
The number of random hops of a mobile particle to nearest neighbor sites per unit time is $D'$ 
and excluded volume conditions are applied (i.e. hop attempts to occupied sites are rejected). 
The corresponding tracer diffusion coefficient 
of an isolated particle in two dimensions is $D=D'/4$. 

For each value of the horizontal coordinate $x$,
the top (bottom) solid particle at the $-$ ($+$) interface can detach with rate $Q\epsilon^n$,
where $n$ is the number of nearest neighbors and $\epsilon < 1$; 
thus, weakly bonded solid particles detach with higher probability than strongly bonded ones. 
Solid particles that detach from the solid become mobile particles. 
A mobile particle  that has at least one nearest neighbor
with the $-$ interface and is at the lowest possible position
above the bottom solid may attach and become a solid particle.
With this rule, no overhang of the $-$ interface can be formed.
Hence, our description of the interfaces enter into the class of Solid-On-Solid (SOS) models.
A similar rule is applied on the $+$ interface.
The attachment rate is $Q$.
Hence, in contrast to detachment, attachment does not depend
on the local configuration of the interface. 

The parameter $\epsilon$ is linked to the bond energy $J$
\begin{align}
    \epsilon=\exp[-J/k_BT], 
\end{align}
We can therefore rewrite the detachment rate as $Q\epsilon^n=Q\exp[-nJ/k_BT]$.
Recalling that the attachment rate is $Q$, we recover the usual SOS
bond-breaking model (and its mapping to the Ising model)~\cite{Saito1996,Gagliardi2022} with bond 
energy $J$ and equilibrium concentration 
$$
c_{\mathrm{eq}}=\exp[-2J/k_BT]=\epsilon^2\, .
$$
In addition, the line stiffness of SOS one-dimensional edges reads~\cite{Saito1996}
\begin{align}
    \tilde \gamma = \frac{k_B T}{2\Omega^{1/2}} \big( \varepsilon^{-1/4} - \varepsilon^{1/4} \big)^2\, .
    \label{eq:stiffness}
\end{align}
Furthermore, the attachment-detachment kinetic constant is~\cite{Gagliardi2022} 
$$
\nu=Q\,.
$$

\subsection{Regimes observed in simulations}

\begin{figure}
\centering
\includegraphics[width=1\linewidth]{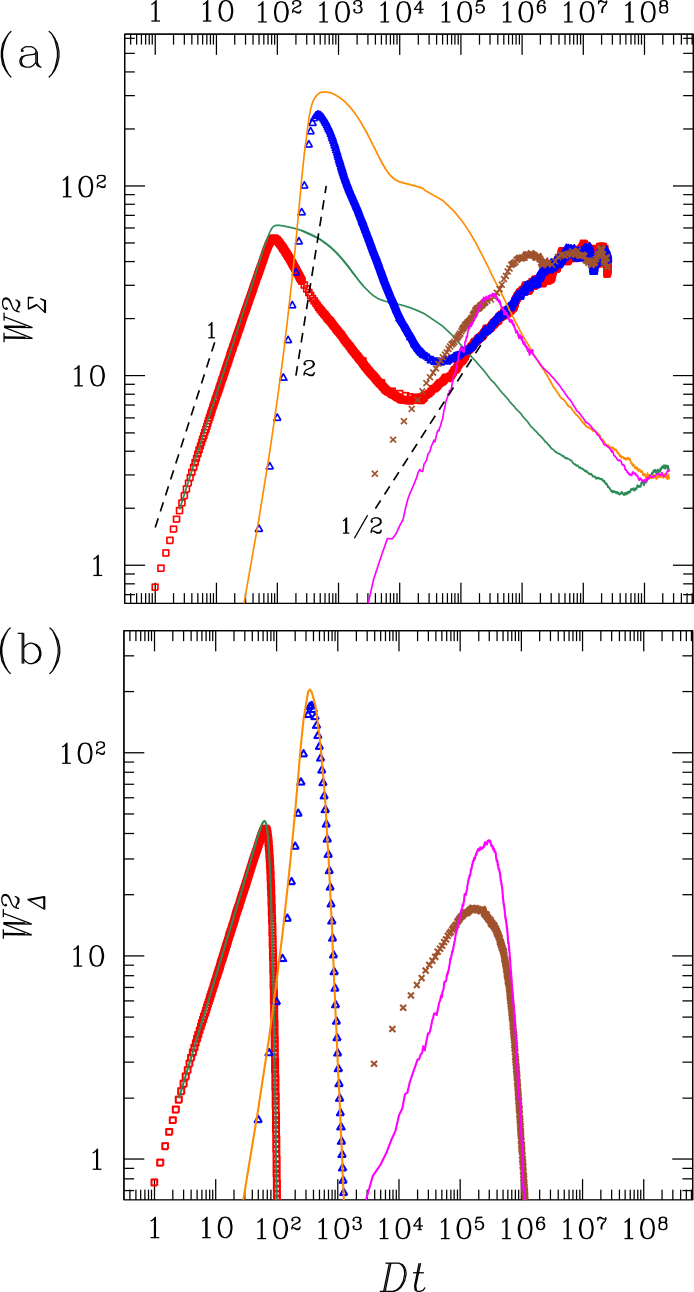}
\caption{
Temporal evolution of the roughness in KMC simulations. 
(a) $W_\Sigma^2$ and (b) $W_\Delta^2$ as a function of $Dt$ for several flux-diffusion ratios:
$F/D=40$ [$\epsilon=0.1$ (red), $\epsilon=0.01$ (green)];
$F/D=4\times{10}^{-3}$ [$\epsilon=0.1$ (blue), $\epsilon=0.01$ (orange)];
$F/D=4\times{10}^{-6}$ [$\epsilon=0.1$ (brown), $\epsilon=0.01$ (magenta)].
The other parameters are $Q/D=0.4$, $d_0=32$, and $L=512$.
The KMC simulation data of these graphs are from Ref.~\cite{Reis2022}.
}
\label{fig:roughness_of_time_KMC}
\end{figure}

Simulations with $L = 512$ and $d_0=32$ are reported
in \cref{fig:roughness_of_time_KMC} for $\epsilon=10^{-2}$ and $10^{-1}$.
Using \cref{eq:stiffness}, these values of $\epsilon$ correspond respectively to $\Gamma=4.05$ 
used in all figures of \cref{sec:Langevin_model},
and to $\Gamma=0.74$.
The general scenario for the evolution of the roughness
is seen to be in qualitative agreement with the results of the Langevin model.
Indeed, for slow growth, we obtain a roughness $W^2_\Sigma$ that increases 
monotonically with time, and then saturates at long times at an equilibrium value.
In contrast, faster growth leads to the formation of a peak of roughness.
Furthermore, the distance-roughness $W^2_\Delta$ follows the 
roughness $W^2_\Sigma$ at short times and then decreases sharply
during the collision as in the Langevin model.
The overall scenario is seen clearly for $\epsilon=10^{-1}$,
but KMC simulations are to slow to reach the asymptotic
equilibrium roughness when $\epsilon=10^{-2}$.

In all plots, the dashed lines are guides to the eye
for a power-law behavior of the squared roughness with the indicated exponent.
If the exponent before the maximum exceeds $2$ during some time interval, the system is considered 
to be unstable. Indeed, the amplitudes  can grow faster than linearly
in the presence of a Mullins-Sekerka instability.
In contrast, when the exponent is near $1/2$, we recover the quasi-equilibrium EW growth, 
which is discussed in Sec.\ref{s:EW_Langevin}.
In Fig.~\ref{Figure_Stability_Diagram_with_shapes}, this simple criterion based on the exponent 
of the power-law before the peak in the Langevin model is shown in orange and red triangles.
This criterion is seen to lead qualitatively to the same three regions as the criterion
based on the evolution of $R_\Sigma(t)$ discussed in Section~\ref{s:classification_Langevin}
(examples of comparison of power laws with 
the results of the Langevin model are reported in Appendix in Fig.~\ref{Figure_exponents_regimes_Langevin}).

\begin{figure}
\centering
\includegraphics[width=1\linewidth]{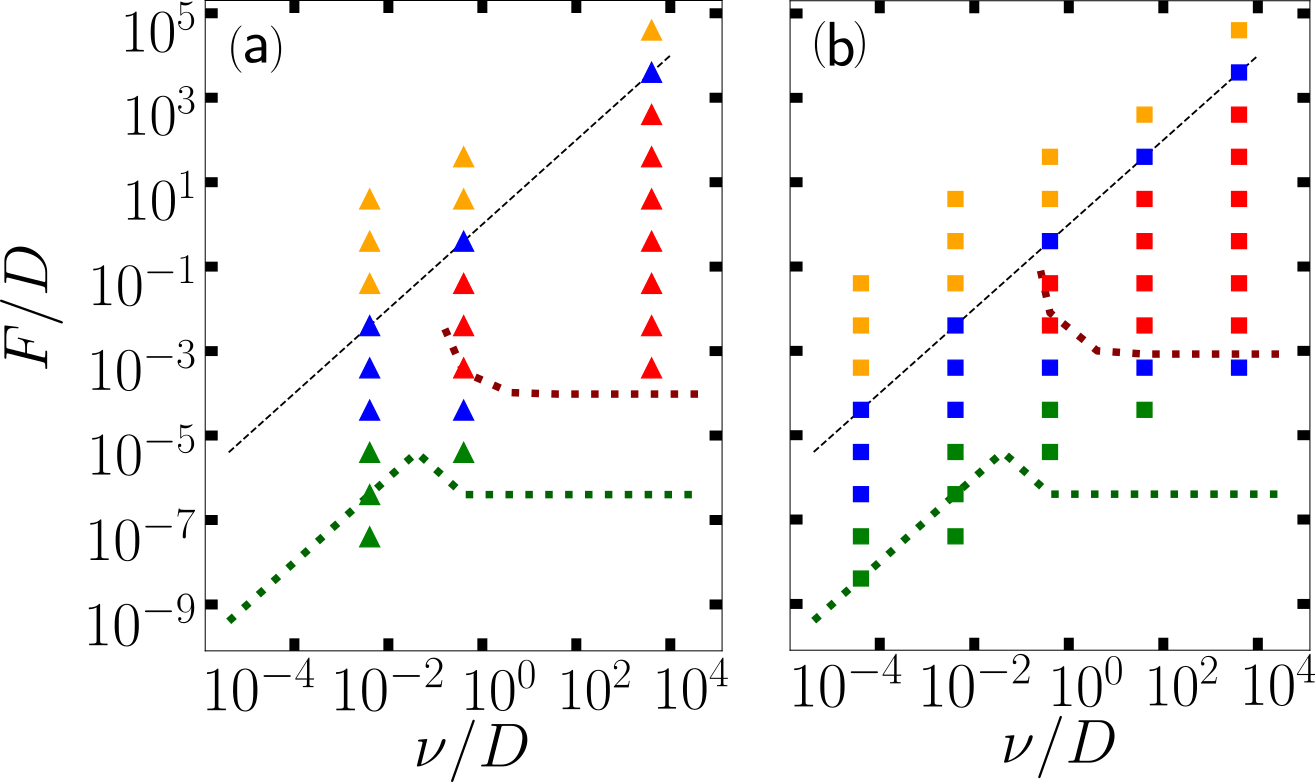}
\caption{Phase diagram from KMC and comparison with the Langevin model.
The symbols represent the regimes identified in KMC simulations.
The dotted lines represent the predictions of the Langevin model.
The dashed line indicates the condition $F/Q=1$.
(a) $h_0=32$ and (b) $h_0=16$. Other parameters: $\epsilon=10^{-1}$, $L=512$. 
The KMC simulation data of these graphs are from Ref.~\cite{Reis2022}.
} \label{phase_diagram_KMC}
\end{figure}

Based on this classification with exponents, 
a phase diagram is also drawn for KMC simulations with $\epsilon=10^{-1}$ in \cref{phase_diagram_KMC}.
The KMC simulations lead to the same three regions as the phase diagram~\cref{Figure_Stability_Diagram_with_shapes},
with qualitative but not precise quantitative agreement.
However for KMC simulations, we also observe
a different region for high fluxes, which corresponds
to a situation where almost all sites on the substrate in the gap
between the two edges are covered by atoms.
In this high-coverage regime, 
the exponent before the maximum is near $1$,
and the roughening  regime is therefore classified as being random deposition (RD)\cite{Barabasi1995}
as already discussed in Sec.~\ref{s:RD_Langevin}.

This regime of high coverage
is obtained when the coverage  close to the edges
becomes of the order of $1$. 
Using the quasistatic expression of the concentration
at the edges \cref{Eqt_c_pm} we obtain a condition for being
in the low-coverage regime $\Omega c\ll 1$, leading to
$
\Omega F\ll {\nu}/{\bar{h}_0^{(0)}(0)}
$.
This condition is qualitatively 
correct, but quantitatively not in agreement with the 
transition to the high-coverage regime 
in \cref{phase_diagram_KMC}.
Indeed, the quasistatic concentration only provides
an upper bound for the possible value of the 
coverage at the edge. A simple condition comparing deposition
and attachment $F/Q<1$ is seen to be 
in better agreement with the simulations,
suggesting as expected, that strong non-quasistatic
effects come to the fore where the concentration is not small.

\subsection{EW regime at low coverage}

In the regime of low coverages, the EW scaling
can be observed. Using \cref{eq:EW_scaling,eq:stiffness},
we obtain
\begin{align}
    W_\Sigma^2 = \frac{4}{\pi^{1/2}} \frac{\varepsilon}{ \varepsilon^{-1/4} - \varepsilon^{1/4}} (Qt)^{1/2}\, .
    \label{eq:KMC_EW}
\end{align}
This expression is valid in the limit $\bar{h}_0^{(0)}Q/D\ll 1$,
which corresponds to our KMC simulations.

\begin{figure}
\centering
\includegraphics[width=\linewidth]{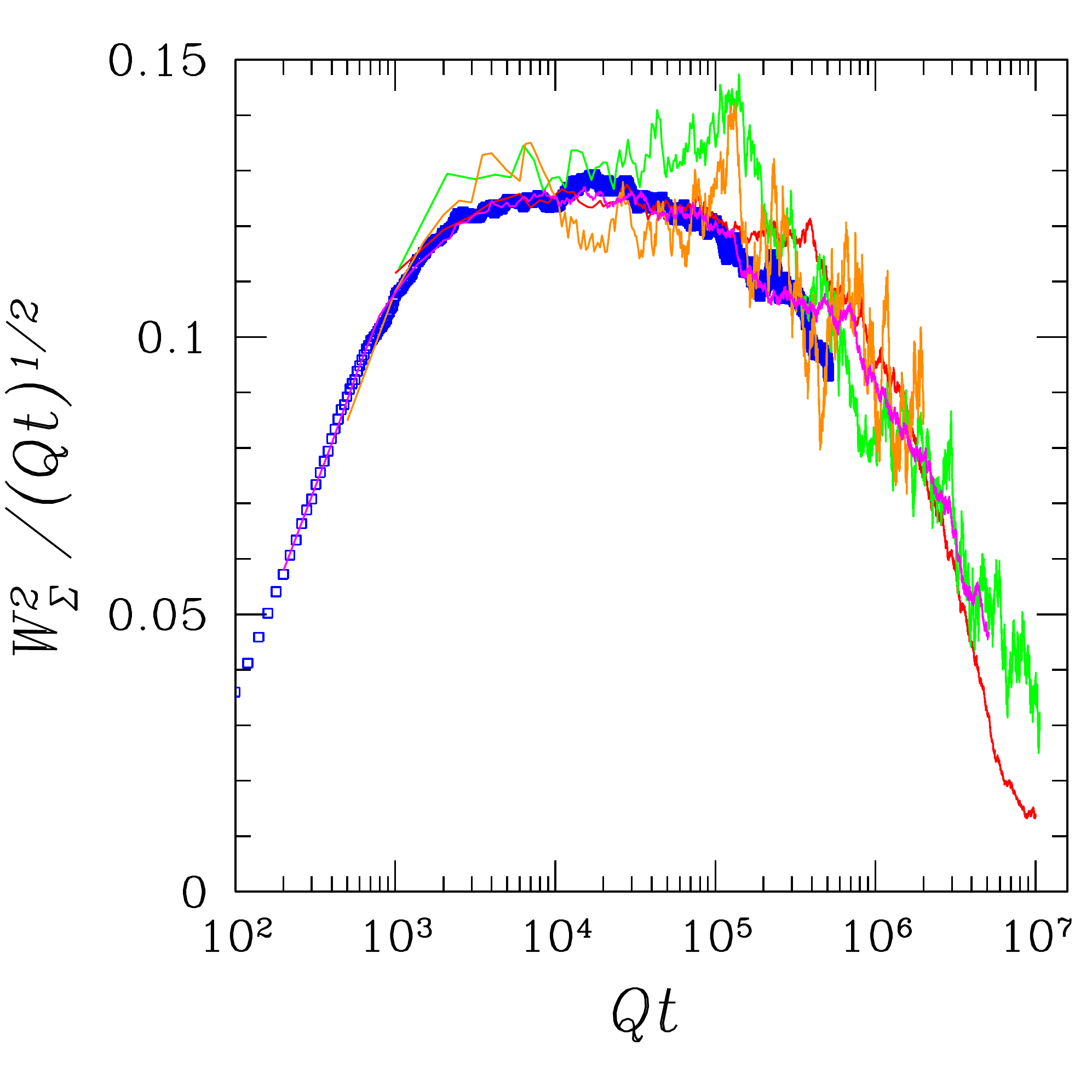}
\caption{Evaluation of the prefactor in the Edwards-Wilkinson scaling of the roughness.
In all cases, $\epsilon=0.1$ and $L=512$. Other parameters:
$(F/D,\nu/D,d_0)=(4\times10^{-10},4\times10^{-4},64$) blue squares;
$(4\times10^{-9},4\times10^{-3},32)$ red curve;
$(4\times10^{-10},4\times10^{-3},32)$ green curve;
$(4\times10^{-10},4\times10^{-3},32)$ green curve;
$(4\times10^{-9},4\times10^{-3},64)$ orange curve;
$(4\times10^{-10},4\times10^{-3},64)$ magenta curve.
} \label{EW_prefactor}
\end{figure}

In \cref{EW_prefactor}, the 
ratio $W_\Sigma^2/(Qt)^{1/2}$ is seen to reach $0.12\pm0.02$
for a wide range of parameters.
Using \cref{eq:KMC_EW} with $\epsilon=10^{-1}$, we find $W_\Sigma^2/(Qt)^{1/2}\approx 0.18$.
Once again, the predictions are providing the correct order of magnitude,
but do not reach precise quantitative agreement.
This difference between the KMC simulations and the predictions of the Langevin model could be caused
by the inaccuracy of the equilibrium line stiffness \cref{eq:stiffness}
in non-equilibrium growth conditions, as discussed in 
other models e.g. in Refs.~\cite{Caflisch1999,Politi1996}.

\subsection{Discussion}

Some features of the KMC simulations
are more delicate to compare to the Langevin model.
For example, the comparison of the short time behaviors is difficult.
A first reason for this difficulty is that  
the short time behavior $W^2\sim t$ analyzed in \cref{s:RD_Langevin}
crucially depends on the details of the atomic cutoff.
In addition, the KMC simulations were started
with no atom in the gap.
The initial build-up of the concentration up to
the quasistatic profile \cref{Eqt_Concentration_parabolic,Eqt_c_pm}
is not described in the Langevin model. In contrast, the Langevin model assumes
an initial condition that starts with the quasistatic profile.
Hence, we have not tried to compare the short time behavior
in KMC and Langevin model.

Another difference is the quantitative value of the
equilibrium asymptotic roughness at long times.
Indeed, the Langevin model assumes that the 
free energy of the grain boundary is simply
the sum of the free energies of the two edges.
This assumption discards the entropic cost
of bringing the two interfaces close to each other.
Such a reduction of the entropy leads to
a larger line tension, and therefore to a larger stiffness
of the grain boundary. This increase
of the stiffness in turn leads to a decrease of the equilibrium 
roughness. A quantitative analysis
of this effect is reported in \cite{Reis2022}.

Despite these limitations in the comparison with KMC simulations,
the Langevin model is seen to be able to 
recover the important and generic features of grain boundary formation,
i.e. the possible non-monotonic behavior of the roughness with a peak and a minimum,
and the generic competition between statistical fluctuations and instabilities
for the production of roughness.

\section{Conclusion}

In summary, we have presented a Langevin model
which aims at describing the formation of grain boundaries
in 2D materials. Previous development of Langevin
models has already proved useful to describe asymptotic power-law
behaviors of the roughness
of one-dimensional interfaces such as atomic steps~\cite{Misbah2010}
during growth or at equilibrium.
However, the formation of grain boundaries is a challenging problem because it
involves transient phenomena occurring in finite time.
A first advance in the modeling of grain boundaries~\cite{Reis2018}
has allowed one to model the collision of two interfaces
in the presence of short-range interactions.
The addition of long-range interactions and fluctuations
associated with the diffusion of growth units
allows one to get closer to existing experimental conditions, but
leads to additional technical difficulties.
Indeed, fluctuations emerge as perturbations of a time-dependent 
reference state that describes the closing of the gap 
between two 2D materials governed by the Zeno effect.

The Langevin model presented in this paper aims at tackling this challenging problem.
This model allows one to identify prominent features
in the formation of grain boundaries
associated to statistical fluctuations and instabilities.
We find that fast-enough growth is accompanied
with a non-monotonous behaviour of the roughness
as a function of time, with a peak and a minimum before
the slow relaxation of the grain boundary after collision towards its equilibrium roughness.
The competition between diffusion-limited instabilities
and statistical noise for the production of the roughness
gives rise to distinctive regimes that can be 
summarized in a phase diagram.
A quantitative comparison of these results with 
KMC simulations reported in Ref.~\cite{Reis2022} is encouraging and 
suggests that our results could serve as
a guide to understand the effect of various
physical parameters in experiments.


\begin{appendix}
\section{Deposition noise}
\label{Appendix_Deposition_Noise}

The goal of this Appendix \ref{Appendix_Deposition_Noise} is to determine 
the amplitudes of the out-of-equilibrium fluctuations in Eq. (\ref{Eqt_Out-of-eq_Autocorrelation-function}). 
We consider a discrete one-dimensional model with two edges growing towards each other 
via the deposition of particles in continuous time. 
We then determine the amplitude of the Langevin forces in a continuum model that
are consistent with the 1D discrete model.

\subsection{One-dimensional lattice model}
\begin{figure}[h]
\centering
\includegraphics[width=\linewidth]{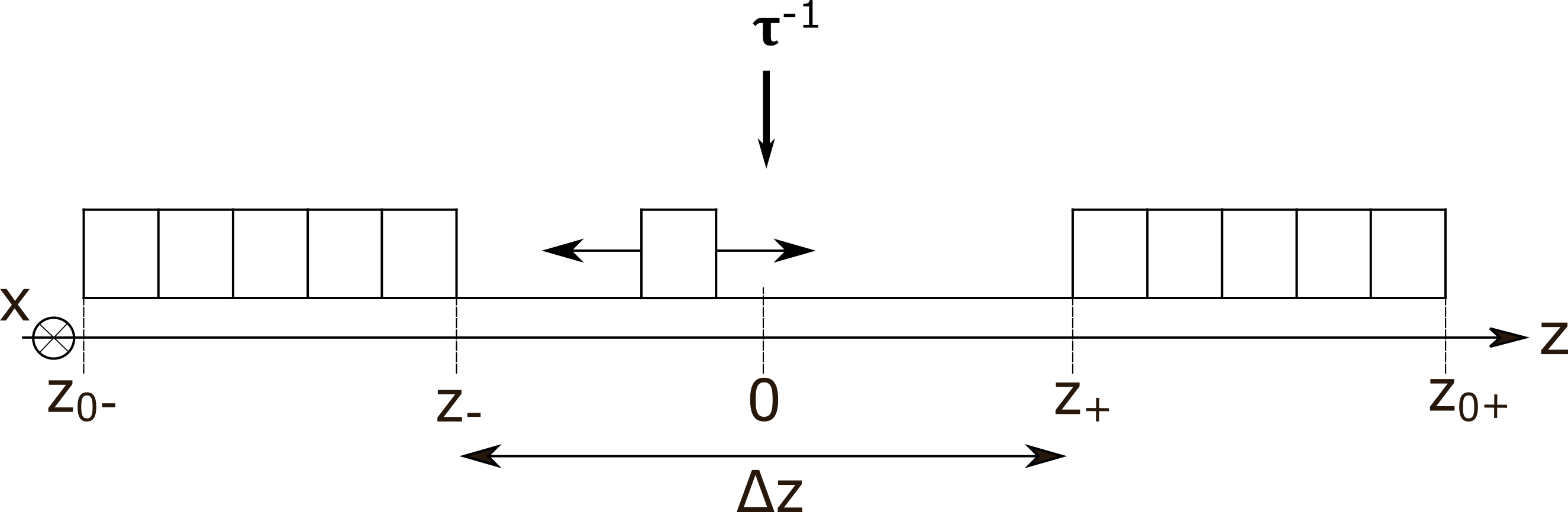}\caption{Two flat bidimensional edges represented as two facing lines as the information along the transversal axis is irrelevant.} \label{Figure_Schema_Appendix}
\end{figure}

The 1D model is composed of two edges defined by their position $z_+$ and $z_-$ along the $z$-axis,
as described on Fig.~\ref{Figure_Schema_Appendix}. 
Deposited particles diffuse along $z$, and attach to one of the edges.
We consider the limit of low deposition flux, where each deposited particle attaches
to one of the edges before another particle has landed. 
Since particles are deposited at a random position in the gap, the deposition-diffusion-attachment process leads to 
random incorporation of particles to the left or to the right with a probability $p=1/2$. 
In the spirit of the quasistatic approximation in the main text, and 
discarding the possibility of slow attachment kinetics, we assume that the process of
diffusion and attachment is instantaneous.

Since we focus on the noise related to freshly landed atoms,
we assume that particles never detach from the edges. Thus,
only one of the two following deposition events can occur:
\begin{align}
    z_- &\rightarrow z_- + 1, \\
    z_+ &\rightarrow z_+ - 1 .
\end{align}
These two events can be written in terms of the in-phase and out-of-phase modes
\begin{align}
\Delta z = z_+ - z_- &\rightarrow \Delta z - 1 , \label{Eqt_Deposition-events_Sigma}\\
\Sigma z = z_+ + z_- &\rightarrow \Sigma z \pm 1. \label{Eqt_Deposition-events_Delta}
\end{align}
The number of empty sites is equal to $\Delta z$. 
We denote $m_+$ as the number of atoms deposited on the $+$ side, 
$m_-$ on the $-$ side, and $\Sigma m = m_+ + m_-$ the total number of deposited atoms. 
We assume a left-right symmetric initial condition with 
$z_{0+}=-z_{0-} = \Delta z_0/2$, and $\Sigma z_0 = 0$,
where $\Delta z_0$ is the initial number of free sites. 
We therefore have $z_- = z_{0-} + m_-$ and $z_+ = z_{0+} - m_+$. 
This leads to a simple bijective relation between $\Delta z$ and $\Sigma m $, and between $\Sigma z$ and $\Delta m$:
\begin{align}
\Delta z &= \Delta z_0 - \Sigma m, \label{Eqt_bijection1}\\
\Sigma z &= \Sigma z_0 - \Delta m = -\Delta m \label{Eqt_bijection2}.
\end{align}

Each site in the gap between the two edges can be filled with a rate $\tau^{-1}$. 
The evolution of $\Delta z$ depends only on the number of deposition events,
and does not depend on the fact that particles are attached to the left or to the right.
Hence, the evolution of $\Delta z$ can be derived from a direct analogy
to a decay process, where a number $\Delta z$ of independent empty sites can be filled by a particle
with a rate $\tau^{-1}$. 
Let $w(t)$ be the probability for a given site to be empty up to a time $t$
\begin{equation} 
\mathrm{d}w(t) = - w(t) \dfrac{\mathrm{d}t}{\tau} \quad \Leftrightarrow \quad w(t) = e^{-\frac{t}{\tau}} .
\end{equation} 
The probability for $\Delta z$ sites to be empty at a given time $t$ is 
\begin{equation} 
P(\Delta z, t)=\left(\begin{array}{l}
\Delta z_{0} \\
\Delta z
\end{array}\right) w^{\Delta z}\left(1-w\right)^{\Delta z_{0}-\Delta z} ,
\end{equation}
which leads to the average number of free sites
\begin{align}
\left\langle\Delta z\right\rangle_t &= \sum_{\Delta z}\left(\begin{array}{l}
\Delta z_{0} \\
\Delta z
\end{array}\right) w^{\Delta z}\left(1-w\right)^{\Delta z_{0}- \Delta z} \Delta z, \nonumber\\
\left\langle\Delta z\right\rangle_t &= w \Delta z_{0} = \Delta z_0 \ e^{-\frac{t}{\tau}} . \label{Eqt_DeltaZ_average}
\end{align}
One retrieves the Zeno effect with the exponential decrease of the gap as in Eq.~(\ref{Eqt_Interface_height}). 
The second moment of $\Delta z$ is given by
\begin{align}
\left\langle\Delta z^2\right\rangle_t &= \sum_{\Delta z}\left(\begin{array}{l}
\Delta z_{0} \\
\Delta z
\end{array}\right) w^{\Delta z}\left(1-w\right)^{\Delta z_{0}- \Delta z} (\Delta z)^2, \nonumber\\ 
 &= w(1-w) \Delta z_0 + \Delta z_{0}^2 w^2 , \label{Eqt_DeltaZ_Raw-Moment_2}
\end{align}
which leads to the following variance
\begin{equation}\label{Eqt_DeltaZ_variance}
V_{\Delta} = \left\langle\Delta z^2\right\rangle_t - \left\langle \Delta z \right\rangle_t^2 
= \Delta z_0 \  e^{-\frac{t}{\tau}} \left(1-e^{-\frac{t}{\tau}}\right).
\end{equation}

In the following, we write the properties of $\Sigma z$
as a function of $\Delta z$.
Since $\Delta z$ monotonously decreases with time, one can directly
switch from expected values at fixed $t$ 
to expected values at fixed $\Delta z$. 
The probability of $\Sigma z$ given a value of $\Delta z$ is denoted $P(\Sigma z|\Delta z)$, 
and the expectation value for a given $\Delta z$ 
is denoted as $\left\langle \ \, \ \right\rangle_{\Delta z}$.
We start with the law of total probability
\begin{equation}
    P(\Sigma z,t) = \sum_{\Delta z=0}^{\Delta z_0} P(\Sigma z|\Delta z) P(\Delta z,t) .
\end{equation}
The n-th moment of $\Sigma z$ at $t$ is written as
\begin{align}\label{Eqt_Moment_orderN_SigmaZ}
    \left\langle \left(\Sigma z\right)^n \right\rangle_{t} &= \sum_{\Sigma z} \left(\Sigma z\right)^n P(\Sigma z,t) 
    \nonumber \\
    &= \sum_{\Delta z=0}^{\Delta z_0} P(\Delta z,t)  \left\langle \left(\Sigma z\right)^n \right\rangle_{\Delta z}  .
\end{align}
where
\begin{equation}
    \left\langle \left(\Sigma z\right)^n \right\rangle_{\Delta z} = \sum_{\Sigma z} \left(\Sigma z\right)^n P(\Sigma z|\Delta z).
\end{equation}

Since $\Sigma z = \Sigma m - 2m_+$, we will evaluate the moments of $m_+$ 
\begin{equation}\label{Eqt_Nth_moment_m+}
\left\langle m_{+}^n\right\rangle_{\Delta z} = \sum_{m_+} m_+ ^n P(m_+|\Delta z) .
\end{equation}
The probability of $m_+$ particles attached to + side among $\Sigma m$ particles deposited is
\begin{equation}
P(m_+|\Delta z) = P(m_+|\Sigma m) = \frac{1}{2^{\Sigma m}}\left(\begin{array}{c}
\Sigma m \\
m_{+}
\end{array}\right).
\end{equation}
One obtains from (\ref{Eqt_Nth_moment_m+}) for n=1 and n=2:
\begin{align}
\left\langle m_{+}\right\rangle_{\Delta z} &= \sum_{m_+} m_+  P(m_+|\Delta z) = \frac{1}{2} \Sigma m, \\
\left\langle m_{+}^{2}\right\rangle_{\Delta z} &= \sum_{m_+} m_+ ^2 P(m_+|\Delta z) = \frac{1}{4} \Sigma m + \frac{1}{4}(\Sigma m)^{2},
\end{align}
which finally leads to 
\begin{align}
&\left\langle \Sigma z \right\rangle_{\Delta z} = \Sigma m - 2\left\langle m_{+}\right\rangle_{\Delta z}=0, \label{Eqt_SigmaZ_average}\\
&\langle \left(\Sigma z\right)^2 \rangle_{\Delta z} = 
\langle \left( \Sigma m - 2m_+ \right)^2 \rangle_{\Delta z} = \Sigma m. \label{Eqt_SigmaZ_variance}
\end{align}
From \cref{Eqt_Moment_orderN_SigmaZ}, the average vanishes 
\begin{align}
    \left\langle \Sigma z \right\rangle_{t} = 0. 
    \label{Eqt_average_Sigma_z_t}
\end{align}
Moreover, the variance is obtained by inserting \ref{Eqt_SigmaZ_variance} using
\cref{Eqt_bijection1} into \cref{Eqt_Moment_orderN_SigmaZ}
\begin{align}
V_{\Sigma} &= \left\langle \left(\Sigma z\right)^2 \right\rangle_{t} 
= \sum_{\Delta z=0}^{\Delta z_0}  P(\Delta z,t) 
\left\langle \left(\Sigma z\right)^2 \right\rangle_{\Delta z}
\nonumber \\
&= \Delta z_0 \left( 1 - e^{-\frac{t}{\tau}}\right) .
\label{Eqt_SigmaZ_variance_discrete}
\end{align}

\subsection{Langevin model}

We now design continuum Langevin equations which are consistent with (\ref{Eqt_DeltaZ_variance}) and (\ref{Eqt_SigmaZ_variance_discrete}). 
The position along the $x$-axis parallel to the edge is here explicitly expressed by 
the discrete index $m$. We assume that the process $\Delta z_m$ and $\Sigma z_m$ 
at a given $m$ is independent from the others. This means physically
that we assume that attachment occurs at the same coordinate $x$ as
the deposition event.
The decrease of the distance between the two edges $\Delta z$ 
is taken to be proportional to the deposition rate as in (\ref{Eqt_DeltaZ_average}):
\begin{align}
\partial_t \Delta z_{m} &= -\dfrac{1}{\tau} \Delta z_{m} + \Tilde{\varphi}_{\Delta,m}(t), \label{Eqt_Sigma_Z_TimeEvolution} \\
\partial_t \Sigma z_{m} &= \Tilde{\varphi}_{\Sigma,m}(t). \label{Eqt_Delta_Z_TimeEvolution}
\end{align}
The Langevin forces have zero average 
$\left\langle \Tilde{\varphi}_{\Delta,m}(t) \right\rangle = \left\langle \Tilde{\varphi}_{\Sigma,m}(t) \right\rangle = 0$. 
They are also uncorrelated in time, and their amplitudes are defined as
\begin{equation}\label{Eqt_Phi_sigma&delta_delta_autocorrelation}
\left\langle \Tilde{\varphi}_{i,m}(t_1) \Tilde{\varphi}_{j,m'}(t_2) \right\rangle = 2 \Tilde{D}_{i,m}(t_1) \delta(t_1 - t_2) \delta_{m,m'} \delta_{i,j} \quad ,
\end{equation}
where $i$ and $j$ are either $\Delta$ or $\Sigma$,
where $\delta_{\mathrm{n,n'}}$ is the Kronecker delta symbol. 
The equations (\ref{Eqt_Sigma_Z_TimeEvolution}) and (\ref{Eqt_Delta_Z_TimeEvolution}) are solved as
\begin{align}
\Delta z_{m} (t) &= \Delta z_{0,m} e^{-t/\tau} + \int_{0}^{t} \mathrm{d}t_1 \ \Tilde{\varphi}_{\Delta,m}(t_1) e^{\frac{1}{\tau}(t_1 - t)}, \\
\Sigma z_{m} (t) &= \int_{0}^{t} \mathrm{d}t_1 \ \Tilde{\varphi}_{\Sigma,m}(t_1) e^{\frac{1}{\tau}(t_1 - t)} .
\end{align}
 This leads to $\left\langle \Delta z_{m} \right\rangle_t = \Delta z_{0,m} e^{-t/\tau}$ and $\left\langle \Sigma z_{m} \right\rangle_t = 0$,
  in agreement with \cref{Eqt_DeltaZ_average,Eqt_average_Sigma_z_t}.
  The variances in the Langevin model read
\begin{align}
V_{\Delta} &= \left\langle \left(\Delta z_{m}\right)^2 \right\rangle_t - \left\langle \Delta z_{m} \right\rangle_t^2 = \int_{0}^{t} \mathrm{d}t_1 \ 2\Tilde{D}_{\Delta,m}  e^{\frac{2}{\tau}(t_1 - t)}, \\
V_{\Sigma} &= \left\langle \left(\Sigma z_{m}\right)^2 \right\rangle_t = \int_{0}^{t} \mathrm{d}t_1 \ 2\Tilde{D}_{\Sigma,m}(t_1).
\end{align}
We then impose the expression of $\Tilde{D}_{\Delta,m}$ and $\Tilde{D}_{\Sigma,m}$ to 
obtain agreement with (\ref{Eqt_DeltaZ_variance}) and (\ref{Eqt_SigmaZ_variance_discrete}):
\begin{equation}\label{Eqt_Intensity}
2\Tilde{D}_{\Delta,m}(t) = 2\Tilde{D}_{\Sigma,m}(t) = \dfrac{1}{\tau} \left\langle \Delta z_{m} \right\rangle_t  .  
\end{equation}
Indeed, using these expressions, we find
\begin{align}
V_{\Delta} 
&= \int_{0}^{t} \mathrm{d}t_1 \ \dfrac{1}{\tau} \left\langle \Delta z_{m} \right\rangle_{t_1} 
= \Delta z_{0,m} \, e^{\frac{-t}{\tau}} \left( 1 - e^{-\frac{t}{\tau}}\right), \label{Eqt_2nd-raw-moment_Delta} \\
V_{\Sigma} &= \int_{0}^{t} \mathrm{d}t_1 \ \dfrac{1}{\tau} \left\langle \Delta z_{m} \right\rangle_{t_1} = \Delta z_{0,m} \left( 1 - e^{-\frac{t}{\tau}}\right) .\label{Eqt_2nd-raw-moment_Sigma}
\end{align}

We now take the continuum limit. Multiplying
Eqs.~(\ref{Eqt_Sigma_Z_TimeEvolution},\ref{Eqt_Delta_Z_TimeEvolution}) by the atomic length $a$,
letting $a\rightarrow 0$,
and using $a\Delta z_m\rightarrow\Delta h(x,t)$, 
$a\Sigma z_m\rightarrow\Sigma h(x,t)$
and $a\Tilde{\varphi}_{i,m}(t) \rightarrow \varphi_{i}(x,t)$ leads to
\begin{align}
\partial_t \Delta h(x,t) &= -\dfrac{1}{\tau} \Delta h(x,t) + \varphi_{\Delta}(x,t), \label{Eqt_Rho_Delta_Z_TimeEvolution}\\
\partial_t \Sigma h(x,t) &= \varphi_{\Sigma}(x,t), \label{Eqt_Rho_Sigma_Z_TimeEvolution}
\end{align}
and using $\delta_{m,m'}\xrightarrow[]{} a\ \delta(x-x')$, we find
\begin{equation}
\left\langle \varphi_{i}(x,t) \varphi_{j}(x',t') \right\rangle = A_{i}(x,t) \delta(x-x') \delta(t-t') (2\pi)^2 \delta_{i,j} .
\end{equation}
where
\begin{align}
A_{\Sigma}(x,t) = 2 \Tilde{D}_{\Sigma}(t) a^3 \, ,
\nonumber \\
A_{\Delta}(x,t)  = 2 \Tilde{D}_{\Delta}(t) a^3 \, .
\end{align}
Finally, using \cref{Eqt_Intensity} and the relations 
$1/\tau=\Omega F$ and $a\left\langle \Delta z_{m} \right\rangle_t\rightarrow 2{\bar{h}}^{(0)}(t)$ ,
we obtain
\begin{equation}
A_{\Delta}(x,t) = A_{\Sigma}(x,t)  = 2 \Omega^2 F {\bar{h}}^{(0)}(t),
\end{equation}
which is identical to (\ref{Eqt_Out-of-eq_Autocorrelation-function}).

\section{Close-to-equilibrium roughening}
\label{a:EW}

In this appendix,
we provide a derivation of the evolution of the roughness
in the limit of small incoming flux $F\rightarrow 0$.
We start with the evolution equation for the power-spectrum
Eq.(\ref{Eqt_roughness_of_modeQ_2_Sum_TimeDerived}).
Changing variables from $t$ to $\Bar{h}^{(0)}(t)$, we obtain
\begin{align}\label{Equation_spectral_h0_derivative}
     \partial_{\Bar{h}^{(0)}}\langle|\Sigma h^{(1)}_{\mathrm{q}}(t)|^2\rangle 
     =& -2  \frac{\lambda_{\Sigma q}}{\Omega F \Bar{h}^{(0)}} \langle|\Sigma h^{(1)}_{\mathrm{q}}(t)|^2\rangle 
     \nonumber\\ 
     &-\frac{B_{\Sigma q}L}{\Omega F \Bar{h}^{(0)}} - 2\Omega L.
\end{align}
Since long wavelength contributions dominate the roughness at
equilibrium, we expect the roughening process to be
dominated by long wavelength modes close to equilibrium.
As a consequence, we take both limits $F\rightarrow 0$ and $q\rightarrow 0$
in the expressions of $\lambda_{\Sigma q}$ and $B_{\Sigma q}$,
leading to
\begin{equation}\label{Eqt_Rouhness_variable-h0}
    \begin{aligned}
    \Omega F \Bar{h}^{(0)} (\Bar{h}^{(0)} +\frac{D}{\nu}) &\partial_{\Bar{h}^{(0)}} \langle |\Sigma h^{(1)}_{\mathrm{q}}(t)|^2 \rangle = \\
    2\Omega D c_{\text{eq}}^{(0)} \Gamma &q^2 \langle |\Sigma h^{(1)}_{\mathrm{q}}(t)|^2 \rangle - 4\Omega^2 c_{\text{eq}}^{(0)} D L.
    \end{aligned}
\end{equation}
Considering a flat initial condition $\langle |\Sigma h^{(1)}_{\mathrm{q}}(0)|^2 \rangle=0$, the solution of this equation reads
\begin{align}
    \langle |\Sigma h^{(1)}_{\mathrm{q}}(t)|^2 \rangle = \frac{2\Omega L^3}{\Gamma (2\pi n)^2} (1- \text{e}^{-v n^2 }) ,
\end{align}
where
\begin{align}
 v= 2\Omega \Gamma \left(\frac{2\pi}{L}\right)^2\frac{\nu c_{\text{eq}}^{(0)}}{\Omega F}
 \ln\frac{1 + \frac{D}{\nu\Bar{h}^{(0)}(t) }}{1 + \frac{D}{\nu\Bar{h}^{(0)}(0) }} >0 .
\end{align}
The $\Sigma$-roughness is then evaluated as
\begin{align}
\label{aeq:WSigma^2_v}
    \langle W_{\Sigma}^2 \rangle &= \frac{\Omega L}{2\pi^2 \Gamma}\sum_{\mathrm{n}\neq 0}^{} \frac{1}{n^2}(1 - \text{e}^{-v n^2}) .
\end{align}

The $\Sigma$-roughness  exhibits different
behaviors when $v\gg1$ and when $v\ll1$.
 In the limit $v\gg1$, the term $\text{e}^{-v n^2}$ is negligible in Eq.(\ref{aeq:WSigma^2_v}), and 
one obtains the expected asymptotic equilibrium value Eq.(\ref{eq:asymptotic_WSigma^2}).
In the opposite limit $v\ll1$, the sum in Eq.(\ref{aeq:WSigma^2_v}) can be approximated by an integral,
\begin{align}
    \langle W_{\Sigma}^2 \rangle 
    &\approx \frac{\Omega L}{2\pi^2 \Gamma}2\int_1^\infty \mathrm{d}n \frac{1}{n^2}(1 - \text{e}^{-v n^2})
    \nonumber \\
    &\approx \frac{\Omega L}{2\pi^2 \Gamma}2v^{1/2} \int_0^\infty  \mathrm{d}x \frac{1}{x^2}(1 - \text{e}^{-x^2})
    \nonumber \\
    &=\frac{\Omega L}{2\pi^2 \Gamma}2v^{1/2} \pi^{1/2}.
    \label{aeq:W^2_Sigma_small_v}
\end{align}
where $x=vn^2$.

Care should be taken because the relation between $v$ and $t$ is nonlinear and depends 
on the attachment-detachment kinetics. 

\begin{figure}
\centering
\includegraphics[width=\linewidth]{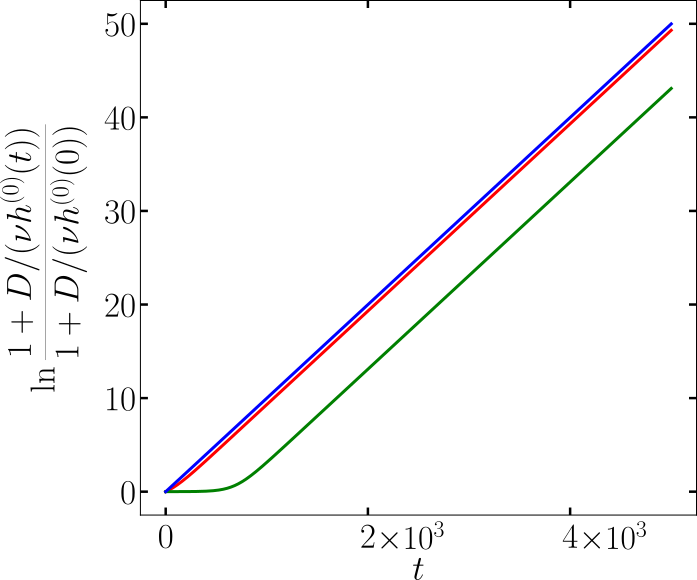}
\caption{The quantity $v\times 4\pi^2F/(2\Gamma\nu c_{\mathrm{eq}}^{0} L^2) $ 
    is plotted as a function of time for: 
    $\nu {\bar{h}}^{(0)}(0)/D=10^{-2}$ (blue), 
    $\nu {\bar{h}}^{(0)}(0)/D=1$ (red), 
    and $\nu {\bar{h}}^{(0)}(0)/D=10^{3}$ (green). 
    The other model parameters are
    $\Omega=1$, $D=10^4/4$, $c_{\mathrm{eq}}^{0}=10^{-2}$, $\Gamma=4.05$, $L=512$, ${\bar{h}}^{(0)}(0)=32$, and $F=10^{-2}$.} \label{Figure_V_function_Appendix}
\end{figure}

However, in general when $\Omega F t\ll 1$, then
\begin{align}
 v\approx 2\Omega \Gamma \left(\frac{2\pi}{L}\right)^2
 \frac{\nu c_{\text{eq}}^{(0)}t}{1 + \nu\Bar{h}^{(0)}(0)/D}.
 \label{aeq:v_short_times}
\end{align}
Thus $v\ll 1$ corresponds to
\begin{align} 
\label{aeq:short_time_v_small}
t\ll \frac{1}{2\Omega c_{\mathrm{eq}}^{(0)}\nu \Gamma} \left(\frac{L}{2\pi}\right)^2\left(1 + \frac{\nu\Bar{h}^{(0)}(0) }{D}\right)
\end{align}
In this regime where both $t\ll1/(\Omega F)$ and the inequality (\ref{aeq:short_time_v_small}) are obeyed,
Eq.(\ref{eq:EW_scaling}) is obtained from the combination
of Eqs.(\ref{aeq:W^2_Sigma_small_v},\ref{aeq:v_short_times}).

\section{Short-time behavior}
\label{Appendix_Short-time_behavior}

An expansion of the evolution equation for the roughness Eq.~(\ref{Eqt_roughness_sum_tot})
to first order in $h^{(0)}(t) - h^{(0)}(0)$ leads to:
\begin{align}
    \langle W_{\Sigma}^2 \rangle &= - \frac{1}{L} \sum^{N}_{\mathrm{n}\neq 0} \big( h^{(0)}(t) - h^{(0)}(0)\big) \Big( \frac{B_{\Sigma q}(\Bar{h}^{(0)})}{\Omega F \Bar{h}^{(0)}} + 2\Omega \Big) \nonumber\\
    &= \big(h^{(0)}(0) - h^{(0)}(t)\big)\frac{1}{L}
    \Big[ \sum^{N}_{\mathrm{n}\neq 0} \frac{B_{\Sigma q}(\Bar{h}^{(0)})}{\Omega F \Bar{h}^{(0)}} + 4\Omega N \Big].
\end{align} 
The first term inside the brackets 
exhibits two limits  for fast and slow attachment-detachment kinetics
\begin{align}
    &\frac{B_{\Sigma q}(\Bar{h}^{(0)})}{\Omega F \Bar{h}^{(0)}} 
    \underset{\nu/D \rightarrow +\infty}{\longrightarrow} \frac{4\Omega c_{\text{eq}}^{(0)}D}{F \Bar{h}^{(0)}}\frac{q }{\tanh k}, \\
   &\frac{B_{\Sigma q}(\Bar{h}^{(0)})}{\Omega F \Bar{h}^{(0)}} \underset{\nu/D \rightarrow 0}{\longrightarrow} \frac{4\Omega c_{\text{eq}}^{(0)}D}{F \Bar{h}^{(0)}}\frac{\nu }{D} = \frac{4\Omega c_{\text{eq}}^{(0)}\nu}{F \Bar{h}^{(0)}}.
\end{align}
To calculate the sum over all modes of the last term, one recalls the discrete-continuum correspondences: 
$q = 2\pi n/L$ and $N=L/2a$, which give $dn=Ldq/2\pi$ and $q= \pi n/N a$. 
Besides, $k = q\Bar{h}^{(0)}$, which leads to $dk = \Bar{h}^{(0)}dq$ and $dn=L/(2\pi\Bar{h}^{(0)})dk$. 
For $2\pi\Bar{h}^{(0)}/L \ll 1$, we then have
\begin{align}
\sum_{\mathrm{n}\neq 0}^{|n|\leq N} \frac{k}{\tanh k} &\simeq \frac{L}{\pi\Bar{h}^{(0)}} \int_{\frac{\pi}{L}\Bar{h}^{(0)}}^{\frac{\pi}{a}\Bar{h}^{(0)}} dk \frac{k}{\tanh k} \nonumber\\
&\simeq \frac{L}{\pi\Bar{h}^{(0)}} \int_{0}^{\frac{\pi}{a}\Bar{h}^{(0)}} dk \frac{k}{\tanh k}.
\end{align}
Since $\Bar{h}^{(0)} \gg a$, we have $\tanh{k}\underset{k \gg 1}{\sim} 1$ and
\begin{align}
\sum_{\mathrm{n}\neq 0}^{|n|\leq N} \frac{q}{\tanh k} \simeq \frac{L}{\pi\Bar{h}^{(0)\,2}} \frac{1}{2} \bigg(\frac{\pi\Bar{h}^{(0)}}{a}\bigg)^2 =\frac{L\pi}{2a^2}. 
\end{align}
In the diffusion limited regime $\nu/D\rightarrow +\infty$, we obtain:
\begin{align}
    \langle W_{\Sigma}^2 \rangle &\underset{\nu/D \rightarrow +\infty}{\simeq} \frac{2\Omega^2}{a}\Big( c_{\text{eq}}^{(0)}\frac{\pi D}{a} + F\Bar{h}^{(0)}(0) \Big) t .
\end{align} 
Similarly, for $\nu/D \rightarrow 0$:
\begin{align}\label{Eqt_Random-deposition_Appendix}
    \langle W_{\Sigma}^2 \rangle &\underset{\nu/D \rightarrow 0}{\simeq} \frac{2\Omega^2}{a}\Big( c_{\text{eq}}^{(0)}2\nu + F\Bar{h}^{(0)}(0) \Big) t .
\end{align}

\begin{figure}
\centering
\includegraphics[width=\linewidth]{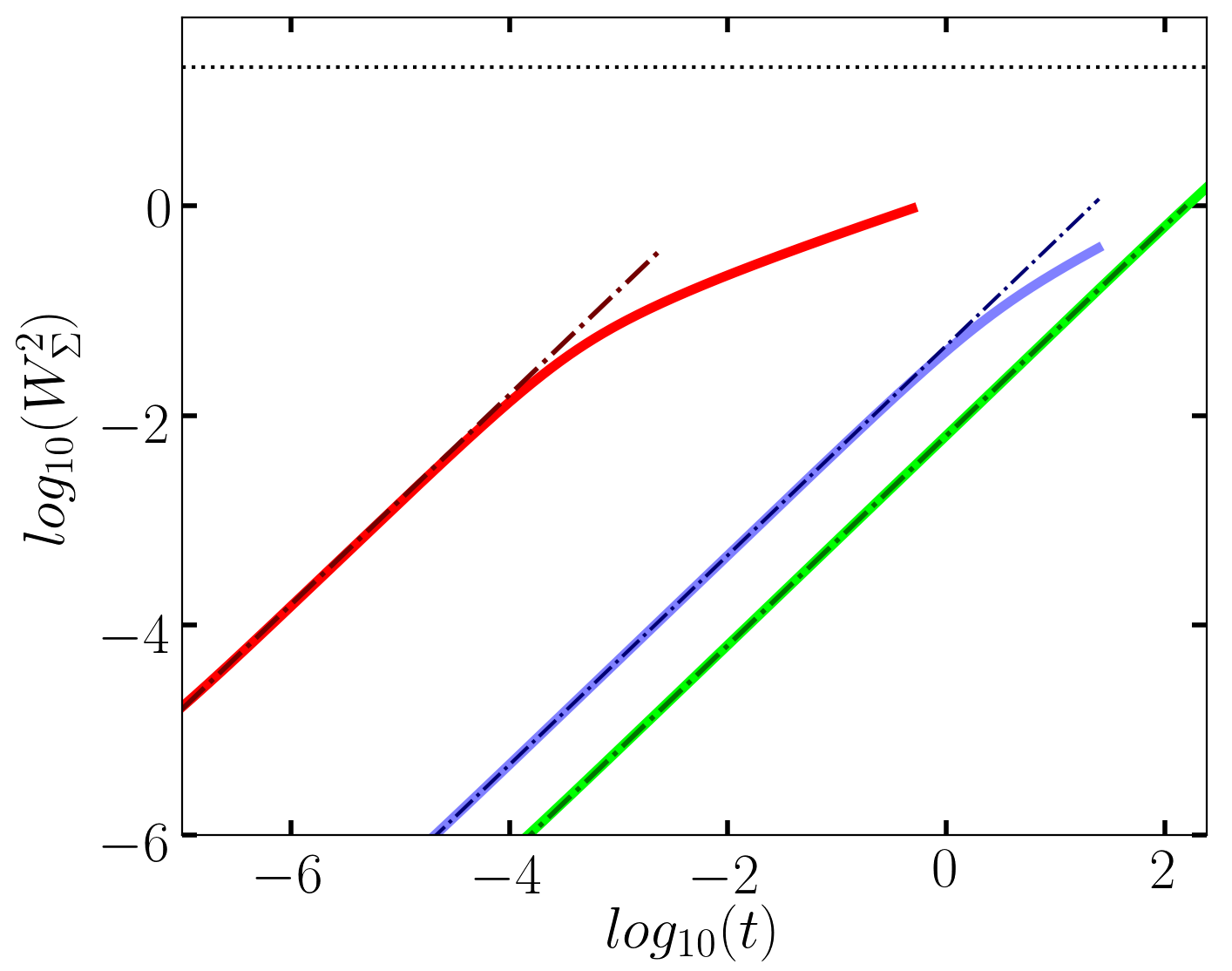}\caption{Roughness at short times for different kinetics. Red $\nu =10^{-3}$, blue $\nu =1$, green $\nu =10^{5}$. The dash-dotted lines are the solution of (\ref{Eqt_Random-deposition_Appendix}). We have used the following model parameters:
    $\Omega=1$, $D=10^4/4$, $c_{\mathrm{eq}}^{0}=10^{-2}$, $\Gamma=4.05$, $L=512$, ${\bar{h}}^{(0)}(0)=32$, $F=10^{-4}$.} \label{Figure_Roughness_Limit_smallF_Appendix}
\end{figure}

\section{Time of the maximum of roughness}
\label{Appendix_Maximum}

In order to investigate the peak of roughness,
we start with (\ref{Equation_spectral_h0_derivative}), where equilibrium fluctuations are neglected:
\begin{align}\label{Equation_spectral_no_equilibrium_fluctuation}
     \partial_{\Bar{h}^{(0)}}\langle|\Sigma h^{(1)}_{\mathrm{q}}(t)|^2\rangle 
     = -2  \frac{\lambda_{\Sigma q}}{\Omega F \Bar{h}^{(0)}} \langle|\Sigma h^{(1)}_{\mathrm{q}}(t)|^2\rangle
     - 2\Omega L.
\end{align}
In the limit of slow attachment-detachment kinetics $\nu \bar{h}^{(0)}/D \ll 1$
and for $k\gg 1$, we obtain
\begin{align}
 \langle|\Sigma h^{(1)}_{\mathrm{q}}(t)|^2\rangle &= \frac{2\Omega L}{1-X_{\mathrm{q}}}\bigg[\bigg(\frac{\bar{h}^{(0)}(t)}{\bar{h}_{0}^{(0)}}\bigg)^{X_{\mathrm{q}}} - \bigg(\frac{\bar{h}^{(0)}(t)}{\bar{h}_{0}^{(0)}}\bigg) \bigg] ,
\end{align}
where
\begin{align}
    X_{\mathrm{q}} = \frac{2 \nu \Gamma c_{\mathrm{eq}}^{0}}{F} q^2 .
\end{align}

For a given mode $q$, the roughness will reach a maximum when $\partial_{\Bar{h}^{(0)}} \langle|\Sigma h^{(1)}_{\mathrm{q}}(t)|^2\rangle = 0$,
i.e. when
\begin{align}
     t = t^q_{\mathrm{peak}} = \frac{-\ln{X_{\mathrm{q}}}}{\Omega F (1-X_{\mathrm{q}})} .
\end{align}
Assuming that the peak is dominated by short-wavelength modes,
we simply consider this condition at the microscopic cutoff $q_{\mathrm{c}}=\pi/a$,
leading to
\begin{align}
    X = X_{\mathrm{q}_{\mathrm{c}}}=\frac{2 \nu \Gamma c_{\mathrm{eq}}^{0}\pi^2}{\Omega F}.
\end{align}
We therefore obtain an estimate of the time of the peak as
\begin{align}
t_{\mathrm{peak}} = \frac{-\ln{X}}{(1-X)}\frac{1}{\Omega F}.
\end{align}

\section{Effective exponents from the Langevin model}

\begin{figure}
\centering
\includegraphics[width=\linewidth]{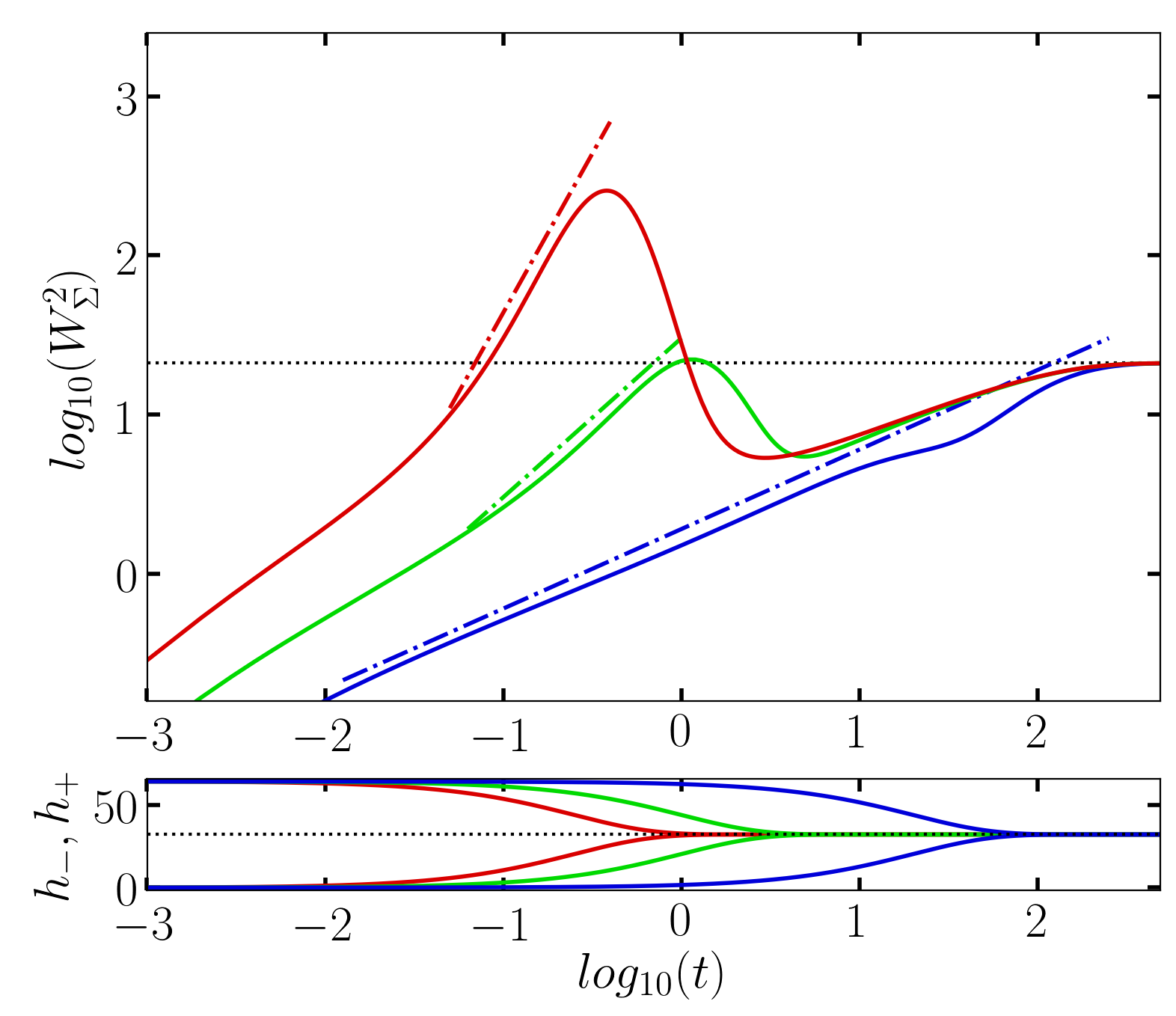}
\caption{
Roughness of the in-phase mode for different values of $F$.
Blue: $F=5.10^{-2}$, green: $F=1$, and red: $F=4$. 
Dashed-dotted lines correspond to power-law scaling. 
Blue: $W_{\Sigma}^2\propto t^{1/2}$,  green: $W_{\Sigma}^2\propto t$, and 
red: $W_{\Sigma}^2\propto t^2$. 
We have used the following model parameters: $\Omega=1$, $D=10^4/4$, $c_{\mathrm{eq}}^{0}=10^{-2}$, $\Gamma=4.05$, $L=512$, ${\bar{h}}^{(0)}(0)=32$, $\nu=10^3$.} 
\label{Figure_exponents_regimes_Langevin}
\end{figure}

\cref{Figure_exponents_regimes_Langevin} illustrates the procedure 
 to extract the effective exponent from the numerical solution of the Langevin model.
These exponents were used to determine the position of the red and 
orange symbols in \cref{Figure_Stability_Diagram_with_shapes}.

\end{appendix}


\bibliography{apssamp}

\end{document}